\documentclass{article}

\usepackage[utf8]{inputenc} % allow utf-8 input
\usepackage[T1]{fontenc}    % use 8-bit T1 fonts
\usepackage{hyperref}       % hyperlinks
\usepackage{url}            % simple URL typesetting
\usepackage{booktabs}       % professional-quality tables
\usepackage{amsfonts}       % blackboard math symbols
\usepackage{nicefrac}       % compact symbols for 1/2, etc.
\usepackage{microtype}      % microtypography
\usepackage{lipsum}
\usepackage{fancyhdr}       % header
\usepackage{graphicx}       % graphics
\graphicspath{{media/}}     % organize your images and other figures under media/ folder
\usepackage{xcolor}
\usepackage{xspace}
\usepackage{subcaption}
\usepackage{multirow}
\usepackage{rotating,tabularx}
\usepackage[numbers]{natbib}
\usepackage[acronym]{glossaries}

\newacronym{DL}{DL}{Deep learning}
\newacronym{CNN}{CNN}{Convolutional Neural Networks}
\newacronym{ASSD}{ASSD}{average symmetric surface distance}
\newacronym{CI}{CI}{Confidence Intervals}

\definecolor{exp1}{HTML}{7CFC00}
\definecolor{exp2}{HTML}{1E90FF}
\definecolor{exp3}{HTML}{FFD700}
\definecolor{exp4}{HTML}{FF8C00}
\definecolor{exp5}{HTML}{696969}
\definecolor{2ndobs}{HTML}{EEAABB}

\newcommand{\mysquare}[1]{\textcolor{#1}{\rule{.24cm}{.24cm}}}
\newcommand{\myexp}[1]{ex\-pe\-ri\-ment~#1~\mysquare{exp#1}}  % \myexp{2} -> \mysquare{exp2} exp 2
\newcommand{\myExp}[1]{Ex\-pe\-ri\-ment~#1~\mysquare{exp#1}}  % \myExp{2} -> \mysquare{exp2} Exp 2
\newcommand{\myEx}[1]{Exp.~#1~\mysquare{exp#1}}  % \myExp{2} -> \mysquare{exp2} Exp 2
\newcommand{\secondobs}{se\-cond~ob\-ser\-ver~\mysquare{2ndobs} }
\newcommand{\secondObs}{Se\-cond~ob\-ser\-ver~\mysquare{2ndobs} }
\newcommand{\dimsddd}[1]{#1$\times$#1$\times$#1}  % \myExp{2} -> \mysquare{exp2} Exp 2
\title{Kidney abnormality segmentation in thorax-abdomen CT scans
}

\author{
  Gabriel Efrain Humpire Mamani$^{1}$, Nikolas Lessmann$^{1}$,\\
  Ernst Th. Scholten$^{1}$, Mathias Prokop$^{1}$,\\
  Colin Jacobs$^{1}$, Bram van Ginneken$^{1,2}$\vspace{0.2cm}\\
  \small $^1$\textit{Diagnostic Image Analysis Group, Radboud University Medical Center,}\\
  \small \textit{Nijmegen, The Netherlands}\vspace{0.05cm}\\
  \small$^2$\textit{Fraunhofer MEVIS, Bremen, Germany}\vspace{0.2cm}\\
  \small \texttt{g.humpiremamani@radboudumc.nl} \\
}
\date{\vspace{-5ex}}

\begin{document}
\maketitle

\begin{abstract}
In this study, we introduce a deep learning approach for segmenting kidney parenchyma and kidney abnormalities to support clinicians in identifying and quantifying renal abnormalities such as cysts, lesions, masses, metastases, and primary tumors. Our end-to-end segmentation method was trained on 215 contrast-enhanced thoracic-abdominal CT scans, with half of these scans containing one or more abnormalities.

We began by implementing our own version of the original 3D U-Net network and incorporated four additional components: an end-to-end multi-resolution approach, a set of task-specific data augmentations, a modified loss function using top-$k$, and spatial dropout. Furthermore, we devised a tailored post-processing strategy. Ablation studies demonstrated that each of the four modifications enhanced kidney abnormality segmentation performance, while three out of four improved kidney parenchyma segmentation. Subsequently, we trained the nnUNet framework on our dataset. By ensembling the optimized 3D U-Net and the nnUNet with our specialized post-processing, we achieved marginally superior results.

Our best-performing model attained Dice scores of 0.965 and 0.947 for segmenting kidney parenchyma in two test sets (20 scans without abnormalities and 30 with abnormalities), outperforming an independent human observer who scored 0.944 and 0.925, respectively. In segmenting kidney abnormalities within the 30 test scans containing them, the top-performing method achieved a Dice score of 0.585, while an independent second human observer reached a score of 0.664, suggesting potential for further improvement in computerized methods.

All training data is available to the research community under a CC-BY 4.0 license on \url{https://doi.org/10.5281/zenodo.8014289}.
\end{abstract}

\section{Introduction}
Kidney cancer is a significant global health issue, ranking as the 12$^{th}$ most deadly cancer in the world, with an estimated 14,700 deaths in 2019 and approximately 73,820 new cases of kidney \& renal pelvis cancer worldwide \cite{Sieg19}. 
With the increasing number of cases, automated tools are needed to assist clinicians in managing this burden.
For instance, by following nephrometry scoring systems~\cite{Kuti09}, automatic kidney tumor segmentation methods may help specialists to detect and get reliable measurements of kidney tumors.

%%%%%%%%%%%%%%%%%%%%%%%% Clinical papers %%%%%%%%%%%%%%%%%%%%%%%%
%%%%%%%%%%%%%%%%%%%%%%%% Computer vision papers %%%%%%%%%%%%%%%%%%%%%%%%
Previous research on kidney segmentation has employed a variety of conventional methods such as region growing \cite{Lin06,Kim04a}, active shape models \cite{Lu07}, active contours \cite{Ling09,Skal17}, graph cut\cite{Ali07,Yoru18}, level-sets\cite{Turc18,Badu16a}, snakes\cite{Farm10}, random forest\cite{Khal17}, and watersheds\cite{Wiec18}. 
However, to the best of our knowledge, there are only a few methods that focus on segmenting kidney tumors or cysts in the literature.
\citet{Ling09} proposed a semi-automatic method that combines fast marching and active geodesic contours to segment renal tumors.
\citet{Kim04a} used thresholds and histograms to segment the kidneys and applied texture analysis to the kidney parenchyma to find seeds for a region-growing algorithm to perform kidney tumor segmentation.
% Patches of liver tumors were analyzed to reduce false positives.
\citet{Chen10g} proposed a method to predict kidney tumor growth in mm$^2$/day, manually segmenting 
the kidney tumors and using a reaction-diffusion model to predict their growth.
\citet{Kaur19} proposed an iterative segmentation method for renal lesions, which uses spatial image details and distance regularization.

%%%%%%%%%%%%%%%%%%%%%% DEEP LEARNING papers %%%%%%%%%%%%%%%%%%%%%%%%
In recent years, \gls{CNN} have shown to be more effective than traditional methods based on classical computer vision techniques and machine learning.
Their ability to learn directly from raw data has led to their widespread use in segmenting organs and structures in different modalities.
For instance, \citet{Zhen17} used an AlexNet-based method to localize the kidneys to define a seed for an active shape model algorithm to segment the kidneys in patients with either abdominal surgery or kidney tumors.
\citet{Shar17} used a network that takes the first 10 layers of the VGG-16 network and upsampled them in a decoder fashion to segment the kidneys of patients with renal insufficiency.
Encoder-decoder networks such us 2D U-Net \cite{Ronn15} and 3D U-Net \cite{Cice16} proved to be robust to tackle medical segmentation tasks in multiple medical imaging segmentation challenges \cite{Bili23,Isen20}.
Variants of these models have been extensively proposed and applied to a wide variety of tasks, including kidney segmentation. 
For instance, \citet{Taha18} segmented the artery, vein, and ureter around the kidneys using a 2D U-Net-like network that allows the deeper layers to influence more to the final prediction.
\citet{Jack18} used a 3D U-Net-like network to segment the kidneys. 
Moreover, several methods used deep learning to segment kidney tumors \cite{Yu19,Yang18}. 
\citet{Yu19} proposed Crossbar-Net, a network that segments kidney and kidney tumors and uses horizontal and vertical patches instead of traditional squared patches. 
The network is divided into sets of sub-networks; a set consists of a sub-network for vertical and another for horizontal patches.
%The vertical and horizontal patches overlap in the middle region.
%This network generates a sequence of two sub-networks (one for vertical and another one for horizontal patches). 
%In total, the method used three sequences of sub-networks. 
%The first sequence was trained using the full data. 
%A new sequence trains the horizontal sub-model with the patches where the vertical sub-model of the previous sequence failed, and vice versa.
%In this way, a new sequence focuses more on the most challenging samples found in a previous sequence.
%The final segmentation takes the majority voting of all the sub-models.
\citet{Yang18} proposed a 3D \gls{CNN} using a pyramid pooling module to segment the kidneys and kidney tumors in abdominal CT angiographic scans.
%The method processes pre-selected kidney ROIs and trains a 3D \gls{CNN} using a pyramid pooling module to get a broader input context.
%Other methods segmented multiple organs, including the kidneys~\cite{Gibs18a,Gibs18c,Hein19,Wang18c}.

The top competitors of the Medical Decathlon~\cite{Isen20} and LiTS challenge~\cite{Chle18,Han17} have achie\-ved the highest performance using cascaded networks. These networks divide the tasks into sub-tasks, with one network per sub-task.
These networks have different fields of view and thus complement each other, resulting in higher performance.
For instance, a first network may segment the liver and the liver tumor as a single structure, aiming to determine the region of interest for the second network; the second network then aims to segment the liver tumor class only.
Similarly, \citet{Blau18} used cascade networks to segment the kidney and kidney cyst in CT scans using a 2D U-Net. 
Their method used heuristics such as a distance transform and HU thresholding to select cyst candidates within the kidney region.
A second (shallow) network classified whether a candidate represented a kidney cyst.
%The second network uses three slices as input (axial, coronal, and sagittal) per cyst candidate. 
Additionally, \citet{Hagh18} used a localization network for pre-processing, which cropped the input for 3D U-Net to segment MRI images of the kidneys. 
In a recent challenge on segmentation of the kidney and kidney tumors on CT \cite{Hell21}, nnUNet \cite{Isen20} was the best performing method.
This method automatically adapts its hyperparameters based on a fingerprint of the data, resulting in optimal performance. Furthermore, it uses 5-fold cross-validation to obtain the final prediction.
%Moreover, despite the potential for promising results, deep learning methods can sometimes misinterpret liver tumors as kidney abnormalities.
%To address this issue, \citet{Kim04a} introduced patches of liver tumors as negative samples to the training set.
%\citet{Shar17} reported a similar problem.

In this study, we propose an automatic method for segmenting the kidney pa\-ren\-chy\-ma and kidney abnormalities in thorax-abdomen CT scans and compare it with the nnUNet. 
We trained our method on 215 thorax-abdomen CT scans and tested on additional 50 scans; the dataset consisted of scans from patients undergoing oncological workup.
The dataset contains patients at different stages of disease and therefore abnormalities can be present in multiple body regions.

\begin{figure}
    \centering
    \includegraphics[width=12cm]{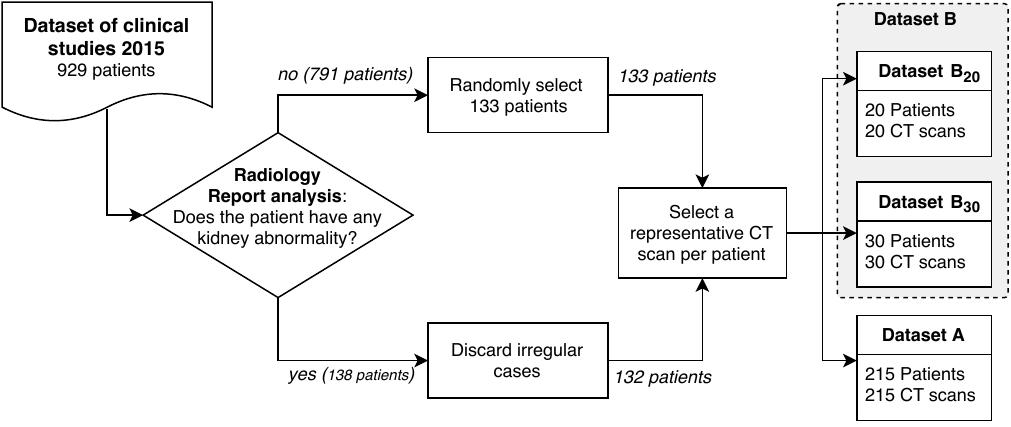}
    \caption{Diagram of the CT scans selection criteria for this study, with dataset A for training and datasets B$_{30}$ and B$_{20}$ for testing (with and without kidney abnormalities respectively).}
    \label{fig:distribution_scans}
\end{figure}
%%%%%%%%%%%%%%%%%%%%%%%%%%%%%%%%%%%%%%%%%%%%%%%%%%%%%%%%%%%%%%%%%%%%%%%%%%%%%%%%%%
%%%%%%%%%%%%%%%%%%%%%%%%%%%%%%%%%%%%%%%%%%%%%%%%%%%%%%%%%%%%%%%%%%%%%%%%%%%%%%%%%%
\section{Materials and Methods}
\subsection{Patient Data}
\label{sec:patient_data}
The dataset used in this study was collected from the Radboud University Medical Center, Nijmegen, the Netherlands.
We randomly retrieved 1905 studies from 929 patients referred by the oncology department in a 12 month period. These patients did not opt-out for use of their data for research, Protected health information was removed from the DICOM data. This retrospective study was approved by the medical-ethical review board of the hospital. 
CT scanners from two manufacturers were used to acquire the CT scans: Toshiba (Aquilion One) and Siemens (Sensation 16, Sensation 64, and Somatom Definition AS).
The reconstruction kernels were FC09, FC09-H, B30f, B30fs, and I30f.
The slice thickness ranged from 0.5 to 3 millimeters, 90\% of them between 1 and 2 mm.
Severe abnormalities throughout the body are present in this dataset resulting from disseminated disease, surgery, chemotherapy, radiotherapy, etc. 

We selected a subset to perform our experiments; the procedure is summarized in Figure \ref{fig:distribution_scans}.
We analyzed the radiology reports per study to intentionally select potential cases that contain kidney abnormalities such as cysts, lesions, masses, metastases, and tumors. In Dutch: \textit{((`cyste' OR `cysten'), (`laesie' OR `lesies'), `massa', (`metastase' OR `metastasen'), and `tumor')}.
Our selection criteria selected studies where the radiology report mentioned in the same sentence the kidneys \textit{('nier' OR 'nieren' NO 'bijnier')} and any kidney abnormalities.
Furthermore, only one clinical study per patient was selected to get a large variety of anatomies for the segmentation task.
In case multiple studies for the same patient were found, we selected 
%the study with most keywords found in the radiology report. 
%Our selection criteria for multiple studies per patient picked the patient 
the study with the earliest acquisition date.

We employed a radiology report analysis to curate a dataset of 138 clinical studies from 138 patients with kidney abnormalities, including cysts, lesions, masses, metastases, or tumors. 
We excluded six patients with unusual anatomy, three patients who had received kidney transplants, two patients with kidneys of irregular size, and one patient with a horseshoe kidney.
The inclusion and exclusion criteria gave us 132 cases for analysis, which were then balanced with additional 133 random patient studies without kidney abnormalities, for a total of 265 CT scans from 265 patients.
The patient cohort contains 56\% males; the average age was 60 years, and the age ranged from 22 to 84.
We divided this set into 215 CT scans for training (dataset A) and 50 for testing (dataset B). The test set was further subdivided, with 60\% (30/50) containing abnormalities (dataset B$_{30}$) and the remaining 40\% (20/50) devoid of abnormalities (dataset B$_{20}$). 
The distribution of the five types of abnormalities (tumors, cysts, masses, lesions, and metastases) was proportional among the 30 cases in dataset B$_{30}$ (six cases per abnormality), which were randomly selected.

In the test set, two and six patients had undergone left and right nephrectomy, respectively, while the training set included seventeen and eighteen patients who had undergone left and right nephrectomy, respectively.

\begin{figure}
    \centering
    \begin{subfigure}[b]{0.32\textwidth}
		\includegraphics[width=3.8cm]{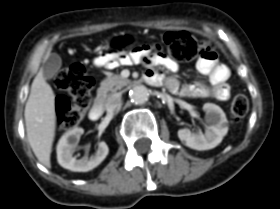}
        \caption{Input CT scan}
    \end{subfigure}
    \begin{subfigure}[b]{0.32\textwidth}
		\includegraphics[width=3.8cm]{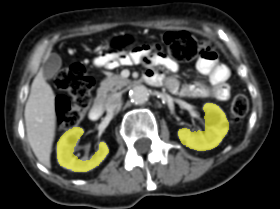}
        \caption{Annotations format 1}
        \label{fig:formats-f1}
    \end{subfigure}
    \begin{subfigure}[b]{0.32\textwidth}
		\includegraphics[width=3.8cm]{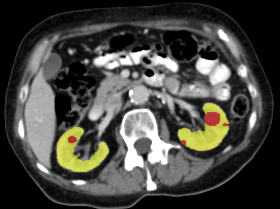}
        \caption{Annotations format 2}
        \label{fig:formats-f2}
    \end{subfigure}
    \caption{Example illustrating the different annotation formats. Each subfigure shows the same axial section, with overlays depicting the annotations: (a) shows the axial CT section. (b) shows the annotations in format 1: parenchyma and kidney abnormalities as a single structure (yellow overlay). (c) shows the annotations in format 2: parenchyma (yellow overlay) and kidney abnormalities (red overlay) as different structures. All images have a window center of 60 HU and a window width of 360 HU.}
    \label{fig:formats}
\end{figure}

\subsection{Annotation procedure}
Four medical students manually segmented the kidney's parenchyma and kidney abnormalities. They were trained by an experienced radiologist (EthS) and consulted the radiologist whenever needed throughout the annotation process. 
Adhering to a standardized protocol, the medical students annotated the kidney parenchyma as the region composed of the renal cortex, renal medulla, and renal pyramid.
The renal hilum, collecting system, and (major and minor) calyces were excluded as much as possible from the kidney parenchyma annotations.
We grouped cysts, lesions, masses, metastases, and tumors connected to the kidney parenchyma as kidney abnormalities. 
The protocol excluded cases with abnormalities in the collecting system.

\begin{figure}[!ht]
%\centering
\begin{subfigure}[b]{6cm}
    \includegraphics[width=5.8cm]{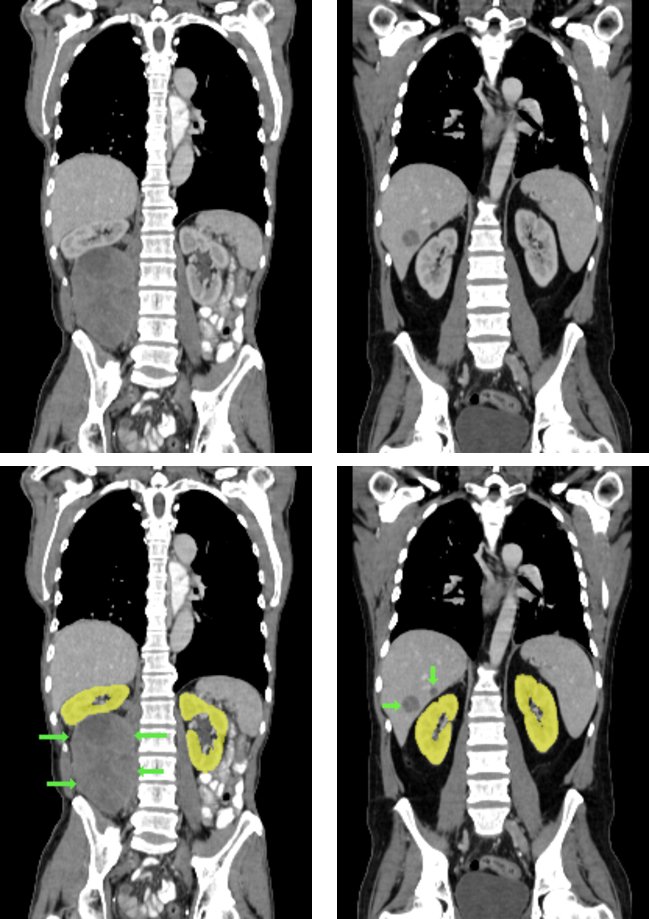}
    \caption{Patients without kidney abnormalities.}
    \label{subfig:dataset_noabn}
\end{subfigure}
\begin{subfigure}[b]{6cm}
    \includegraphics[width=5.8cm]{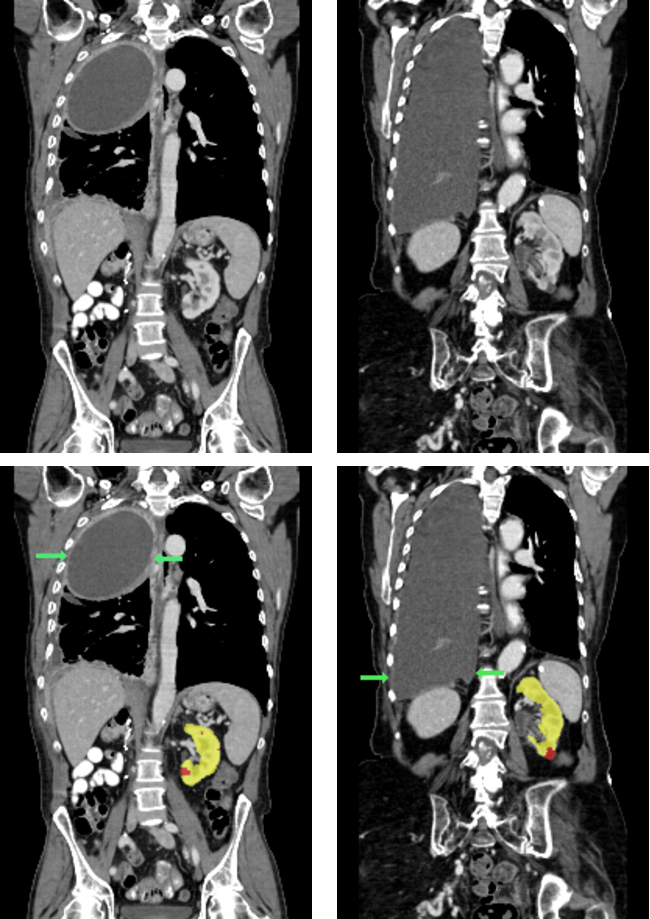}
    \caption{Patients with kidney abnormalities.}
    \label{subfig:dataset_abn}
\end{subfigure}
  \caption{Four examples of CT scans from the training set (dataset A) showing coronal sections with annotations in format 2 (see Figure \ref{fig:formats-f2}) where yellow and red overlays represent annotations of the parenchyma and kidney abnormalities, respectively. Note that all the patients have anomalies in the body (green arrows in the body), and both cases of (b) have only one kidney and contain kidney abnormalities. All the slices have a window center of 60 HU and a window width of 360 HU.}
  \label{fig:dataset_example}
\end{figure}

Annotators used an in-house tool based on MeVisLab \cite{Heck09} to fully delineate the contours of the structures in 2D orthogonal planes.
Our tool was designed to reduce the manual annotation time by interpolating unannotated contours between two manually delineated contours.
The kidney parenchyma of the training set was annotated using an active learning process, with medical students correcting the kidney parenchyma predictions made by a pre-trained 3D U-Net (it used 50 CT scans from dataset A); the kidney abnormalities were annotated from scratch.
The test set was manually annotated (i.e.\, the contour interpolation option of our tool was disabled) by two medical students. One of these was considered as the reference standard and the other one as the \secondobs.
The latter was the most experienced among the medical students and was not allowed to consult the experienced radiologist during these annotations.
The annotations of the \secondobs served as a benchmark for human performance.
The annotations were initially obtained in the axial plane, followed by corrections in coronal and sagittal planes to keep the annotation consistent in all orthogonal directions.

This study utilized two annotation formats, format 1 and format 2, to store the annotations. 
Format 1 considers the kidney parenchyma and kidney abnormalities as a single class (see Figure \ref{fig:formats-f1}) while format 2 separates them into two classes (see Figure \ref{fig:formats-f2}).

Samples of CT scans from patients included in this study can be seen in Figure \ref{fig:dataset_example}. 
While Figure \ref{subfig:dataset_noabn} depicts patients without kidney abnormalities, it highlights the presence of abnormalities in other parts of the body, such as liver tumors.
Figure \ref{subfig:dataset_abn} shows patients with kidney abnormalities, as well as other abnormalities in the body, such as nephrectomy and collapsed lung.

\subsection{Segmentation network}
We present an end-to-end method for segmenting renal parenchyma and abnormalities in CT scans.
We depict our architecture in Figure \ref{fig:diagram_lowres_extension}. It consists of two segmentation networks, a multi-resolution network for kidney segmentation (annotations in format 1, one voxel represents \dimsddd{4}mm) and a high-resolution network (annotations in format 2, one voxel represents \dimsddd{1}mm).
The multi-resolution network is designed to first provide a rough localization of the kidney by processing a low-resolution version of the CT scan. This defines an ROI for the high-resolution network to refine the segmentation of the kidneys and kidney abnormalities.

\subsubsection{Pre-processing}
The CT scans and annotations were resampled to \dimsddd{1}$mm$ (for high-resolution segmentation using annotations in format 2) and \dimsddd{4}$mm$ (for multi-resolution segmentation using annotations in format 1) resolutions (see Figure \ref{subfig:diagram}).
Scans and annotations were resampled using cubic and nearest-neighbor interpolation, respectively.
We clipped the Hounsfield Units to the range [-500,400].

\subsubsection{Multi-resolution network}
\begin{figure}
    \begin{subfigure}[b]{\textwidth}
        \centering
        \includegraphics[width=12cm]{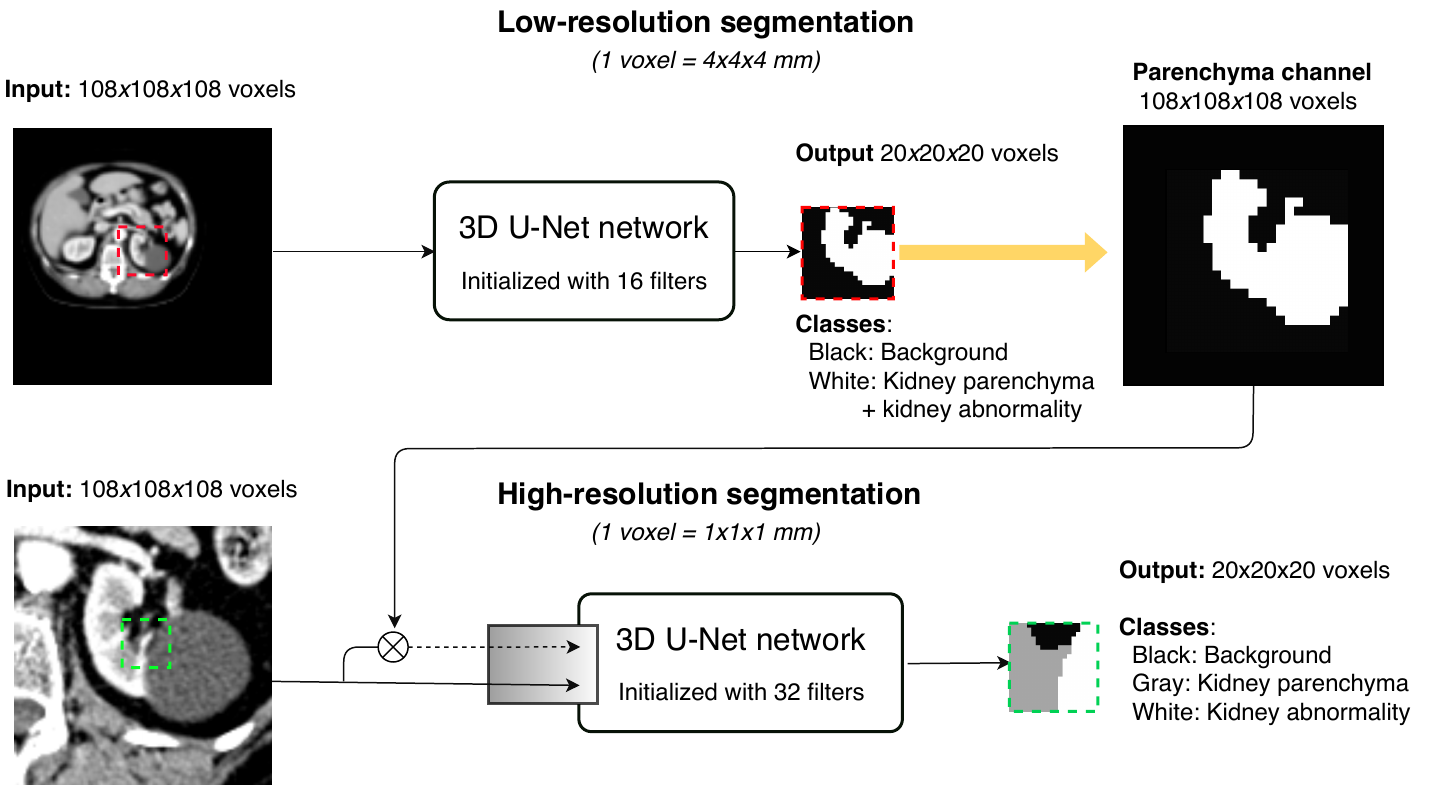}
        \caption{Diagram of the proposed multi-resolution network.}
        \label{subfig:diagram}
    \end{subfigure}\vskip20pt
    \begin{subfigure}[b]{0.6\textwidth}
        \centering
        \includegraphics[width=7cm]{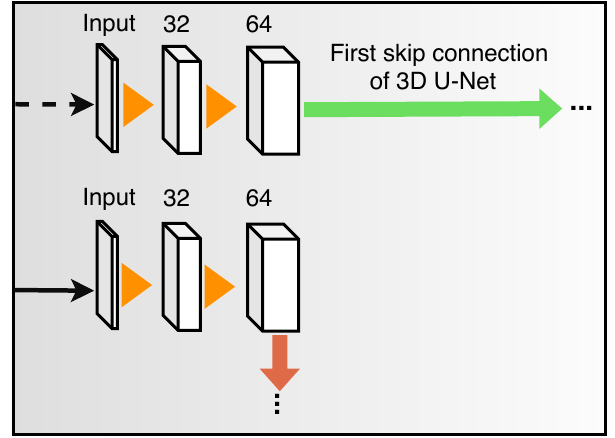}
        \caption{Input of the high-resolution segmentation.}
        \label{subfig:inputmultires}
    \end{subfigure}
    \begin{subfigure}[b]{0.35\textwidth}
        \centering
        \includegraphics[width=4.5cm]{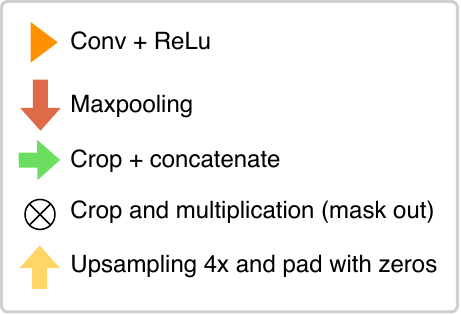}
        \vspace{1.3cm}
    \end{subfigure}
    
    \caption{(a) Diagram of the proposed network. The multi-resolution segmentation network uses a 3D U-Net network initialized with 16 filters. It processes blocks of \dimsddd{108} voxels and outputs the central \dimsddd{20} voxels (represented by the dashed red square). One voxel corresponds to a resolution of \dimsddd{4}mm, giving the network a receptive field of \dimsddd{88} voxels or \dimsddd{352}mm. The kidney parenchyma and kidney abnormalities are considered a single class in the multi-resolution network (see Figure \ref{fig:formats-f1}). The high-resolution segmentation network uses a 3D U-Net architecture initialized with 32 filters, with each voxel representing \dimsddd{1}mm. Its receptive field is \dimsddd{88} mm and it segments the parenchyma and the kidney abnormalities as different classes (see Figure \ref{fig:formats-f2}). (b) Shows how the multi-resolution and the high-resolution networks are connected.}
    \label{fig:diagram_lowres_extension}
\end{figure}

%Traditional cascade networks use two separated networks, one to define the ROI and one for high-resolution segmentation. 
%The disadvantage of cascade networks is that they are not end-to-end (the second \gls{CNN} cannot backpropagate gradients to the first \gls{CNN}). 
We present an end-to-end cascade method for parenchyma and kidney abnormality segmentation.
Unlike traditional cascade networks, which use two separate networks and do not allow for backpropagation, our approach uses a single network composed of two sub-networks.
The first sub-network is a 3D U-Net with 16 filters that performs multi-resolution segmentation and defines an ROI.
This network takes 3D patches of \dimsddd{108} voxels, with each voxel representing \dimsddd{4} mm, as input using annotations in format 1 (kidney parenchyma + kidney abnormalities) and outputs \dimsddd{20} voxels. 
The output is then up-sampled 4 times and padded with zeros to match and mask out the high-resolution input image in millimeters (\dimsddd{108} mm, one voxel represents \dimsddd{1} mm).
The masked-out image serves as an additional input to the second sub-network, the high-resolution segmentation network, which uses a 3D U-Net with 32 filters and serves to fine-segment the kidneys and kidney abnormalities (see Figure \ref{subfig:inputmultires}).
%This branch of filters connects to the skip connection of the high-resolution segmentation network.
Figures \ref{subfig:diagram} and Figure \ref{subfig:inputmultires} illustrate our approach and the connection between the multi-resolution and the high-resolution segmentation network, respectively.

\subsubsection{Data augmentation}
\label{sec:data_augmentation}
Data augmentation was applied randomly to 70\% of the training samples using scaling, rotation, Gaussian blurring, image intensity variation, and elastic deformation.
Up to three of these data augmentation methods were applied randomly to each training sample, to prevent too much data distortion.
When elastic deformation was used, it was only performed in conjunction with Gaussian blurring and image intensity variation.
Interpolation methods of cubic and nearest neighbor were used for CT scans and reference standards, respectively. 
The scaling factor ranged from 0.95 to 1.05, with rotations of up to two planes of -5$^\circ$ to 5$^\circ$ degrees. Gaussian blurring had a sigma range of 0.2 to 1.0, and image intensity variation varied between -20 and 20 HU.
We performed elastic deformation by placing ten control points in a grid, randomly perturbed by up to 5 voxels that were used as input to cubic B-spline interpolation. 

\subsubsection{Spatial dropout}
We applied spatial dropout \citep{Tomp15}, a regularization technique that is different from traditional dropout. Spatial dropout drops feature maps instead of individual neurons to enforce independence among feature maps, encouraging the network to learn more robust and generalizable features. We randomly dropped 10\% of the feature maps per layer.

\subsubsection{Loss function}
The loss function determines how the network's weights are optimized after a forward pass.
In our experiments, we used a combination of weighted categorical cross-entropy and dice loss in the experiments.
\begin{eqnarray}
\mbox{Combined loss} &=&\alpha*diceLoss + \gamma*TopK(weightedCrossentropy)
\end{eqnarray}
where $\alpha=0.3$ and $\gamma=0.7$ were used in all the experiments.
Top-$k$ \cite{Berr18a} sorts the voxel-wise loss in descending order and keeps the top $k\%$ to compute the final mean loss; this approach emulates an online voxel-wise hard-mining per sample.

\subsubsection{Post-processing}
The output of the networks was post-processed to eliminate false positives.
The end-user prediction was reconstructed by stitching together the predictions.
In all the networks, the output was thresholded at 0.5 to get a binary prediction.
The predictions of the multi-resolution network were up-sampled four times and dilated five times to mask out the predictions of the high-resolution segmentation network.
Only the kidney abnormalities that were connected to the kidney parenchyma were kept, to ensure that there were no spurious kidney abnormality candidates outside the kidney region.

\subsubsection{\gls{CNN} Settings}
Due to the large footprint of the network, scans were divided into 3D patches to train the 3D network.
Each training sample consisted of a patch of \dimsddd{108} voxels from the CT scan and a \dimsddd{20} voxel reference standard.
During training, the reference standard patches were sampled every ten voxels in all the orthogonal planes with up to 50\% overlap among surrounding patches.
During inference, the cubes do not overlap.
Patches at the border of the CT scan were mirrored to match the input network size.

The Glorot uniform algorithm \citep{Glor10} was used to initialize the weights of the network.
The weight-map $w$ compensated for the high-class imbalance between the classes.
The background, parenchyma, and kidney abnormality classes had empirically defined weights of 0.05, 0.10, and 0.99, respectively.
We used Adam \citep{King15} as optimization function with learning rate$=0.00001$, $\beta_1=0.9$, and $\beta_2=0.999$. 
The training stopped when the performance on the validation set stopped improving for ten epochs, and the model with the highest average Dice score on the validation set was selected as the optimal model. 

\subsubsection{Implementation of the \gls{CNN}}
The networks were implemented using Keras and TensorFlow as backend in Python 3.6.
The segmentation experiments were executed on a cluster of computers equipped with GTX1080 and GTX1080ti graphics cards, each with 256GB of CPU RAM.

\subsection{Evaluation}
The end-user segmentation obtained by our networks was compared to the reference masks using the Dice score. %, precision, sensitivity, and false negative rate.
\begin{eqnarray}
%\mbox{Dice score} &=&\frac{2 * TP}{2 * TP + FP + FN}  %,\\ % \frac{2|X \cap Y |}{|X|+|Y|}
\mbox{Dice score} &=& \frac{2*volume(X \cap Y )}{volume(X)+volume(Y)}
% \mbox{Precision} &=&\frac{TP}{TP + FP},\\
% \mbox{Sensitivity/Recall/TPR} &=&\frac{TP}{TP + FN},\\
% \mbox{False Negative Rate} &=&\frac{FN}{TP + FN}
\end{eqnarray}
where X is the prediction, and the Y is the reference standard.

\subsection{Ablation study}
\label{sec:ablationstudy}
In this section, we conducted a step-by-step evaluation of the impact of each module (multi-resolution, data augmentation, top-k, and spatial dropout) in our proposed network.
The backbone architecture for this ablation study was the 3D U-Net \cite{Cice16}.
Our experiments setup started with a 3D U-Net, and additional modules were added one by one in subsequent experiments (see the left side of Table \ref{tab:ablation_results}).
In order to evaluate the impact of each module on the network performance, we conducted an ablation study by adding modules to the 3D U-Net backbone architecture one by one.
The baseline network, referred to as \myexp{5}, only used the 3D U-Net initialized with 32 filters and had a single input of \dimsddd{108} voxels with \dimsddd{1} mm per voxel, producing \dimsddd{20} voxels.
The subsequent experiments added the multi-resolution module (\myexp{4}), data augmentation module (\myexp{3}), top-$k$ module (\myexp{2}), and spatial dropout module (\myexp{1}) to the network.
The input and output sizes and formats were consistent across all experiments except \myexp{5}; networks receive two inputs of \dimsddd{108} voxels each, one input of \dimsddd{1} mm and one input of \dimsddd{4} mm per voxel for high-resolution (input of \dimsddd{108} mm using annotation format 2) and multi-resolution segmentation (input of \dimsddd{432} mm using annotation format 1), respectively.
The difference in performance between \myexp{1} (experiment with spatial dropout) and \myexp{2} (experiment without spatial dropout) showed the influence of the spatial dropout module, for example.
As an initial step, we first trained the multi-resolution module independently to reach its optimal sub-model. 
%This is because of the connection between the multi-resolution module masks out ($\otimes$ symbol in Figure \ref{subfig:diagram}) the high-resolution input (\dimsddd{1}$mm$) to serve as a second input to and the 3D U-Net (see Figure \ref{subfig:inputmultires}).
Afterward, we froze the weights of the multi-resolution sub-model, except for the last three layers to allow back-propagation from the high-resolution segmentation network.
All the experiments used 80\% of dataset A for training and 20\% for validation.
Each experiment was trained independently to find the optimal model. 
%The models of \myexp{1}, \myexp{2}, \myexp{3}, \myexp{4}, and \myexp{5} were trained for 25, 16, 19, 19, and 16 epochs, respectively.
The best model from each experiment was evaluated using test sets B$_{20}$ and B$_{30}$.

\subsection{nnUNet}
We conducted experiments with nnUNet \cite{Isen20} to compare its performance with our methods.
Unlike our approach, nnUNet processes CT scans without any preprocessing step, while we resample the CT scans to an isotropic resolution and clip the HU range.
To gain insight about the benefits of ensemble networks, we ensembled nnUNet with our two highest-performing methods, one at a time. 
As nnUNet only uses thresholding as postprocessing, we analyzed the impact of our dedicated postprocessing method on performance.
Note that our postprocessing eliminates false-positive kidney abnormalities that are not attached to the parenchyma.

\begin{figure}[!ht]
    \centering
  \begin{subfigure}[b]{0.49\textwidth}
    \includegraphics[width=5.5cm]{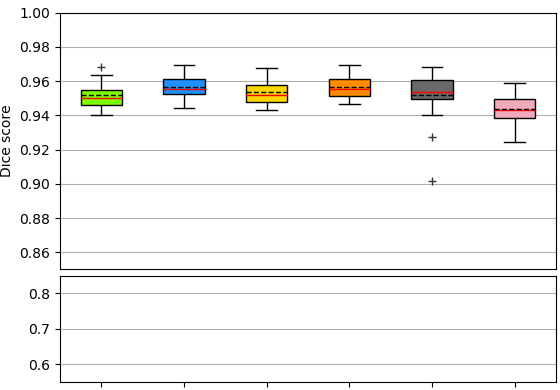}
    \caption{}
    \label{subfig:boxplot_parenchyma_noabn}
  \end{subfigure}
  \hfill\hfill
  \begin{subfigure}[b]{0.49\textwidth}
    \includegraphics[width=5.5cm]{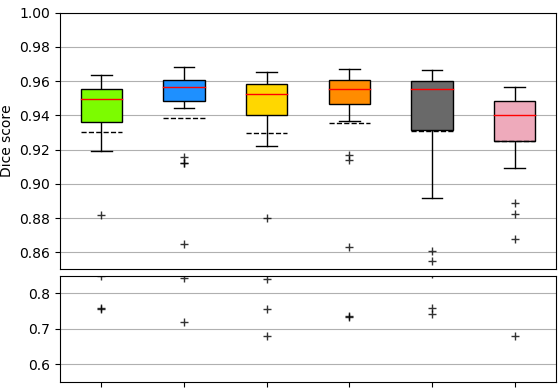}
    \caption{}
    \label{subfig:boxplot_parenchyma_abn}
  \end{subfigure}
  \begin{subfigure}[b]{0.49\textwidth}
    \includegraphics[width=5.5cm]{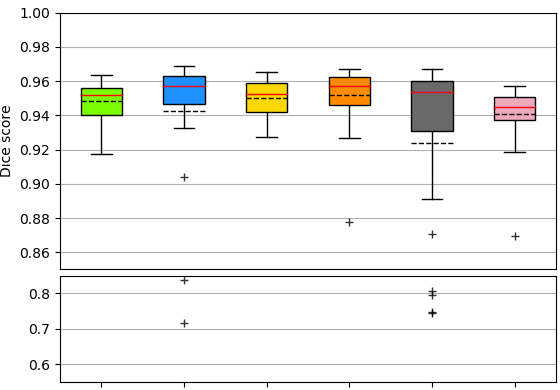}
    \caption{}
    \label{subfig:boxplot_abnparenchyma_abn}
  \end{subfigure}
  \begin{subfigure}[b]{0.49\textwidth}
    \includegraphics[width=5.5cm]{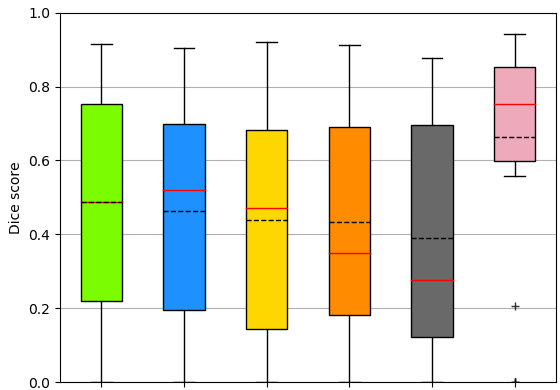}
    \caption{}
    \label{subfig:boxplot_abn_abn}
  \end{subfigure}
  \caption{Performance comparison of our methods and the \secondobs on datasets B$_{20}$ and B$_{30}$ using boxplots. The red and black lines represent the median and the mean, respectively. Boxplot (a) shows results for class parenchyma only on the dataset B$_{20}$ (twenty cases without abnormalities). Boxplot (b) shows results for class parenchyma only on the dataset B$_{30}$ (thirty cases with abnormalities). Boxplot (c) displays the results for class parenchyma plus abnormalities as a single structure on dataset B$_{30}$ (thirty test cases with abnormalities). Boxplot (d) shows results for Class abnormalities only on the dataset B$_{30}$ (thirty cases with abnormalities). Note that the scale in the y-axis is different for boxplot (d). The modules for each experiment are represented by the same color coding as in Table \ref{tab:ablation_results}: \myexp{1}, \myexp{2}, \myexp{3}, \myexp{4}, \myexp{5}, and \secondobs.}
  \label{fig:boxplots}
\end{figure}

\begin{sidewaystable}
%\begin{table}[!ht]
    \scriptsize\centering
    \caption{Summary of performance of our methods and the \secondobs on the test sets (dataset B$_{20}$ and B$_{30}$) using Dice score. The upper part of the table shows Mean and SD, while the bottom part shows 95\% Confidence Intervals. We analyzed the performance per class (parenchyma and kidney abnormalities) and when both classes are combined (parenchyma + kidney abnormalities). An asterisk ($^*$) indicates that a value is significantly better or worse than the value of the reference system (\myexp{1}). The best Dice score values per column are in bold.}
    \label{tab:ablation_results}
    \begin{tabular}{p{3.2cm}cccc}
        \hline
\multirow{5}{*}{\textbf{Experiment}}
    &\multirow{2}{*}{\textbf{Test set: dataset B$_{20}$}}
    %& 
    &\multicolumn{3}{c}{\multirow{2}{*}{\textbf{Test set: dataset B$_{30}$}}}\\
& & & &\\ \cline{2-5} %\cline{2-2}\cline{4-6}
&\multirow{2}{2.2cm}{\centering \textbf{Parenchyma class}} 
     &\multirow{2}{2.2cm}{\centering \textbf{Parenchyma class}}
     &\multirow{2}{2.7cm}{\centering \textbf{Parenchyma + abnormalities class}}
     &\multirow{2}{2.5cm}{\centering \textbf{Abnormalities class}}\\ 
& & & &\\
& (Fig. \ref{subfig:boxplot_parenchyma_noabn}) 
     & (Fig. \ref{subfig:boxplot_parenchyma_abn})
     & (Fig. \ref{subfig:boxplot_abnparenchyma_abn})
     & (Fig. \ref{subfig:boxplot_abn_abn})\\ \hline
\multicolumn{5}{c}{\underline{\textit{Mean $\pm$ SD}}}\\[0.2cm]
\myEx{1} & 0.952$\pm$0.008 & 0.930$\pm$0.053 & 0.948$\pm$0.011 & 0.487$\pm$0.314 \\
\myEx{2} & $^*$0.957$\pm$0.006 & $^*$0.938$\pm$0.051 & $^*$0.942$\pm$0.049 & 0.464$\pm$0.320 \\
\myEx{3} & 0.954$\pm$0.007 & 0.930$\pm$0.064 & 0.950$\pm$0.010 & 0.440$\pm$0.310 \\
\myEx{4} & $^*$0.956$\pm$0.007 & 0.936$\pm$0.058 & $^*$0.952$\pm$0.017 & 0.434$\pm$0.311 \\
\myEx{5} & 0.952$\pm$0.015 & 0.931$\pm$0.057 & 0.924$\pm$0.065 & 0.390$\pm$0.315 \\
\secondObs & $^*$0.944$\pm$0.009 & $^*$0.925$\pm$0.051 & $^*$0.941$\pm$0.017 & \bf0.664$\pm$0.274 \\ \hline
%nnUNet Fold1& $^*$0.960$\pm$0.014 & $^*$0.938$\pm$0.074 & $^*$0.922$\pm$0.120 & 0.483$\pm$0.293 \\
nnUNet& $^*$0.960$\pm$0.015 & $^*$0.940$\pm$0.069 & $^*$0.931$\pm$0.099 & 0.521$\pm$0.303 \\
Ens. nnUNet \myEx{1}& $^*$0.962$\pm$0.007 & $^*$0.941$\pm$0.058 & $^*$0.927$\pm$0.125 & 0.526$\pm$0.305\\
Ens. nnUNet \myEx{2}& $^*$0.964$\pm$0.006 & $^*$0.944$\pm$0.055 & $^*$0.928$\pm$0.129 & 0.507$\pm$0.318 \\ \hline
%nnUNet \textbf{postproc} Fold1& $^*$0.964$\pm$0.006 & $^*$0.940$\pm$0.071 & $^*$0.954$\pm$0.042 & 0.575$\pm$0.284 \\
nnUNet PP& $^*$0.964$\pm$0.006 & $^*$0.941$\pm$0.068 & $^*$0.955$\pm$0.042 & 0.576$\pm$0.290 \\ 
Ens. nnUNet \myEx{1} PP & $^*$0.962$\pm$0.007 & $^*$0.942$\pm$0.058 &\bf $^*$0.960$\pm$0.009 &0.585$\pm$0.293 \\
Ens. nnUNet \myEx{2} PP &\bf $^*$0.965$\pm$0.006 &\bf $^*$0.947$\pm$0.050 & $^*$0.958$\pm$0.032 & 0.566$\pm$0.309 \\
\hline\hline
\multicolumn{5}{c}{\underline{\textit{Confidence Intervals 95\%}}}\\[0.2cm]
\myEx{1}&0.948-0.955& 0.910-0.950&0.945-0.952&0.369-0.604 \\
\myEx{2}&0.954-0.960& 0.920-0.957&0.924-0.961&0.345-0.584 \\
\myEx{3}&0.950-0.957& 0.906-0.954&0.946-0.954&0.324-0.555 \\
\myEx{4}&0.953-0.959& 0.914-0.957&0.945-0.958&0.318-0.550 \\
\myEx{5}&0.945-0.959& 0.910-0.952&0.900-0.948&0.273-0.508 \\
2nd observer&0.939-0.948& 0.906-0.944&0.935-0.947&0.561-0.766 \\ \hline
%nnUNet \textbf{Fold1}& 0.954-0.966 & 0.909-0.961 & 0.874-0.957 & 0.376-0.584 \\
nnUNet& 0.953-0.965 & 0.913-0.961 & 0.892-0.959 & 0.411-0.625 \\
Ens. nnUNet \myEx{1}&  0.959-0.965 & 0.918-0.958 & 0.876-0.959 & 0.416-0.627 \\
Ens. nnUNet \myEx{2}&  0.961-0.967 & 0.923-0.961 & 0.876-0.961 & 0.397-0.613 \\ \hline
%nnUNet postproc \textbf{Fold1}& 0.961-0.967 & 0.911-0.961 & 0.937-0.964 & 0.474-0.670 \\
nnUNet PP& 0.961-0.967 & 0.915-0.962 & 0.938-0.965 & 0.469-0.673 \\
Ens. nnUNet \myEx{1} PP& 0.959-0.966 & 0.918-0.959 & 0.957-0.963 & 0.482-0.687 \\
Ens. nnUNet \myEx{2} PP& 0.962-0.967 & 0.926-0.963 & 0.945-0.965 & 0.454-0.672 \\
        \hline
	\end{tabular}
%\end{table}
\end{sidewaystable}
%%%%%%%%%%%%%%%%%%%%%%%%%%%%%%%%%%%%%%%%%%%%%%%%%%%%%%%%%%%%%%%%%%%%%%%%%%%%%%%%%%
%%%%%%%%%%%%%%%%%%%%%%%%%%%%%%%%%%%%%%%%%%%%%%%%%%%%%%%%%%%%%%%%%%%%%%%%%%%%%%%%%%
\section{Results}
The results of the ablation study conducted on the test sets (dataset B$_{20}$ and B$_{30}$) are shown in Figure \ref{fig:boxplots}. These results are also summarized in Table \ref{tab:ablation_results}, which includes asterisks (*) to indicate statistical significance (P-value $<0.05$) between \myexp{1} and other experiments, as determined by a two-tailed Mann-Whitney U test.
We evaluated the predictions of each experiment per class to show more insights into the results of our experiments.
Furthermore, we combined the prediction of both classes (annotation format 2) as a single structure (annotation format 1) and computed its Dice score; this helps to make our results comparable to methods that reported kidney dice only.

\textbf{\textit{Dataset B$_{30}$:}} The presence of kidney abnormalities characterizes the patients in this dataset (see Figure \ref{subfig:dataset_abn}). 
The results of our experiments on dataset B$_{30}$ are displayed in Figures \ref{subfig:boxplot_abn_abn}, \ref{subfig:boxplot_parenchyma_abn}, and \ref{subfig:boxplot_abnparenchyma_abn}. 
First, we evaluated the performance of the methods in segmenting the kidney abnormalities class only. The results are shown in Figure \ref{subfig:boxplot_abn_abn} and in the column ``Dataset B$_{30}$/Abnormalities class'' of Table \ref{tab:ablation_results}. 
The \secondobs and \myexp{1} achieved the two highest scores, 0.664$\pm$0.274 and 0.487$\pm$0.314, respectively.
\myExp{5} obtained 0.390$\pm$0.315 Dice, the lowest score when segmenting the kidney abnormalities only.
Next, we evaluated the performance of the methods in segmenting the parenchyma class only. The results are shown in Figure \ref{subfig:boxplot_parenchyma_abn} and in the column ``Dataset B$_{30}$/Parenchyma class'' of Table \ref{tab:ablation_results}.
The two highest scores were obtained by \myExp{2} and \myexp{4} with 0.938$\pm$0.051, 0.936$\pm$0.058, respectively, while the \secondobs obtained the lowest score with 0.925$\pm$0.051.
Finally, we evaluated the performance of the methods when segmenting both the parenchyma and the kidney abnormalities class as a single structure (annotation format 1).
The results are shown in Figure \ref{subfig:boxplot_abnparenchyma_abn} and in column ``Dataset B$_{30}$/Parenchyma + abnormalities class'' of Table \ref{tab:ablation_results}.
The two highest scores were achieved by \myExp{4} and \myexp{3} with Dice scores 0.952$\pm$0.017 and 0.950$\pm$0.010, respectively.
\myExp{5} obtained the lowest Dice score with 0.924$\pm$0.065.

\textbf{\textit{Dataset B$_{20}$:}} The patients in this dataset do not present kidney abnormalities, but it is probable that they have other anomalies in the body (see Figure \ref{subfig:dataset_noabn}).
The results on the test set B$_{20}$ are depicted in Figure \ref{subfig:boxplot_parenchyma_noabn} and in Table \ref{tab:ablation_results} under the column ``Dataset B$_{20}$/Parenchyma class''.
\myExp{2} and \myexp{4} obtained the highest Dice scores, 0.957$\pm$0.006 and 0.956$\pm$0.007, respectively.
The \secondobs obtained the lowest Dice score with 0.944$\pm$0.009.

\textbf{\textit{nnUNet:}} 
In our experiments, nnUNet obtained slightly better results in the pa\-ren\-chy\-ma class of datasets B$_{20}$ and B$_{30}$ compared to our experiments, a Dice score of 0.521 $\pm$ 0.303 in the kidney abnormality class, which was higher by +0.034 Dice than our \myexp{1}.
To further analyze the differences between nnUNet and our experiments, we ensembled the predictions of nnUNet with either \myexp{1} or \myexp{2} by averaging their probabilities.
The ensemble nnUNet with \myexp{2} slightly improved the results of nnUNet in the parenchyma class of both datasets but decreased in -0.014 Dice score in the abnormality class, while the ensemble nnUNet with \myexp{1} slightly improved in +0.004 dice score compared to nnUNet in the abnormality class.
The ensemble nnUNet with \myexp{2} performed slightly better than the ensemble nnUNet with \myexp{1} in all classes, except the abnormality class, where the ensemble with \myexp{1} had a Dice score of 0.526 $\pm$ 0.306, and the ensemble with \myexp{2} obtained 0.507 $\pm$ 0.318.
Since nnUNet only uses thresholding for post-processing, we applied our dedicated post-processing to the nnUNet predictions to remove kidney abnormalities that are not attached to the kidney, which resulted in notable improvements of +0.055, +0.059, and +0.059 for nnUNet, ensemble nnUNet with \myexp{1}, and ensemble nnUNet with \myexp{2}, respectively.
As a result, the ensemble nnUNet with \myexp{1} and our dedicated post-processing was the highest-performing experiment in the abnormality class, with a Dice score of 0.585 $\pm$ 0.293. 

%Our results demonstrate that \myexp{1} is our best method for segmenting kidney abnormalities, while \myexp{2} performs slightly better in the parenchyma class.
Table \ref{tab:results_literature} compares our results with other methods published in the literature.
Some of the methods report the Dice scores for the left and right kidneys separately, while others report a single score for both kidneys combined.
To make our results comparable to these methods, we post-processed our predictions to obtain the Dice scores for both the left and right kidneys.

\begin{sidewaystable}
%\begin{table}[!ht]
    \centering
    \scriptsize
    %\tiny
    \caption{Performance comparison between our methods (\myexp{1} and \myexp{2}) and previous work of kidney segmentation and kidney tumor/abnormality segmentation using the mean Dice score as the metric. The methods listed below the line reported the presence of cases with kidney tumors/abnormalities in their datasets. The values marked with $\dag$ are obtained after post-processing the prediction masks from `Both kidneys' column for comparability with other methods.}
    \label{tab:results_literature}
    \hspace{-1cm}
    \begin{tabular}{lccccccl}
            \hline
\multirow{3}{*}{\textbf{Method}}&\multirow{3}{0.7cm}{\textbf{Kidney abnor-malities}}&\multicolumn{3}{c}{\textbf{Kidneys}}& \multirow{3}{0.55cm}{\textbf{Num. test scans}}&\multirow{3}{0.9cm}{\textbf{Deep Learning}}&\multirow{3}{*}{\textbf{Description}}\\ \cline{3-5}

&&\multirow{2}{1cm}{\textbf{Left kidney}}   &\multirow{2}{1cm}{\textbf{Right kidney}} &\multirow{2}{1.1cm}{\textbf{Both kidneys}}&\\
%&&\multirow{2}{0.7cm}{\textbf{Left kidney}}   &\multirow{2}{0.7cm}{\textbf{Right kidney}} &\multirow{2}{0.84cm}{\textbf{Both kidneys}}&\\ % This is when using \tiny
&&&&&&&\\ \hline

\multicolumn{8}{c}{\textit{\underline{Methods that reported cases without kidney abnormalities or did not report them}}}\\
&&&&&&&\\
\citet{Jack18}&N/A&0.860&0.910&-&24&Yes&\\
\citet{Gibs18a,Gibs18c}&N/A&0.930&-&-&\textit{10}&Yes&\textit{9-folds cross-validation}.\\
\citet{Badu16a}&N/A&0.938&0.944&-&20&No&\\
\citet{Hein19}&N/A&0.942&-&-&\textit{10}&Yes&\textit{4-fold cross validation}.\\
\citet{Wang18c}&N/A&0.956&0.954&-&\textit{30}&Yes&\textit{4-fold cross-validation}.\\ 
\citet{Khal17}&N/A&-&-&\textbf{0.973}&\textit{60}&No&\textit{Leave-one-out}.\\ 
\myEx{1}&N/A&0.953$^\dag$&0.951$^\dag$&0.952&20&Yes&Test set: Dataset B$_{20}$, train set: Dataset A.\\
\myEx{2}&N/A&0.960$^\dag$&\textbf{0.955}$^\dag$&0.957&20&Yes&Test set: Dataset B$_{20}$, train set: Dataset A.\\
\secondObs&N/A&0.943$^\dag$&0.945$^\dag$&0.944&20&N/A&Test set: Dataset B$_{20}$, train set: N/A.\\
nnUNet&N/A&0.956$^\dag$&0.959$^\dag$& 0.960&20&Yes&Test set: Dataset B$_{20}$, train set: Dataset A.\\
Ens. nnUNet \myEx{1} PP&N/A&0.962$^\dag$&0.944$^\dag$& 0.962&20&Yes&Test set: Dataset B$_{20}$, train set: Dataset A.\\
Ens. nnUNet \myEx{2} PP&N/A&\textbf{0.964}$^\dag$&0.946$^\dag$& 0.965&20&Yes&Test set: Dataset B$_{20}$, train set: Dataset A.\\
\hline

\multicolumn{8}{c}{\textit{\underline{Methods that reported cases with kidney abnormalities}}}\\
&&&&&&&\\
\citet{Turc18}&-&-&-&0.800&55&No&Polycystic kidneys only.\\
\citet{Shar17}&-&-&-&0.860&81&Yes&Polycystic kidneys.\\
\citet{Skal17}&-&-&-&0.862&10&No&Kidney cancer.\\
\citet{Blau18}&-&0.870&0.870&-&46&Yes&Cysts.\\
\citet{Lin06}&-&0.873&0.886&-&30&No&Two cases with tumor, one with a cyst.\\
\citet{Zhen17}&-&0.890&0.920&-&78&Yes&Kidney tumors.\\
\citet{Wiec18}&-&-&-&0.917&170&No&Cysts and kidney tumors.\\
\citet{Yang18}&0.802&-&-&0.931&50&Yes&Kidney tumors.\\ 
\citet{Yu19}&\textbf{0.913}&-&-&-&36&Yes&Kidney tumors.\\ 
%\citet{Xia18}&-&-&-&-&5&Yes&Kidney tumors\\  
\myEx{1}&0.488&0.949$^\dag$&0.951$^\dag$&0.948&30&Yes&Test set: Dataset B$_{30}$, train set: Dataset A.\\
\myEx{2}&0.464&0.956$^\dag$&0.936$^\dag$&0.942&30&Yes&Test set: Dataset B$_{30}$, train set: Dataset A.\\
\secondObs&0.664&0.939$^\dag$&0.943$^\dag$&0.941&30&N/A&Test set: Dataset B$_{30}$, train set: N/A.\\
nnUNet& 0.521& 0.928$^\dag$&0.951$^\dag$& 0.931&30&Yes&Test set: Dataset B$_{30}$, train set: Dataset A.\\
Ens. nnUNet \myEx{1} PP& 0.585& 0.960$^\dag$&\textbf{0.960}$^\dag$& \textbf{0.960}&30&Yes&Test set: Dataset B$_{30}$, train set: Dataset A.\\
Ens. nnUNet \myEx{2} PP& 0.566& \textbf{0.963}$^\dag$& 0.955$^\dag$& 0.958&30&Yes&Test set: Dataset B$_{30}$, train set: Dataset A.\\
            \hline
    \end{tabular}
%\end{table}
\end{sidewaystable}

%%%%%%%%%%%%%%%%%%%%%%%%%%%%%%%%%%%%%%%%%%%%%%%%%%%%%%%%%%%%%%%%%%%%%%%%%%%%%%%%%%
%%%%%%%%%%%%%%%%%%%%%%%%%%%%%%%%%%%%%%%%%%%%%%%%%%%%%%%%%%%%%%%%%%%%%%%%%%%%%%%%%%
\section{Discussion}
In this paper, we presented an automatic method for the segmentation of the (kidney) parenchyma and kidney abnormalities.
We conducted experiments in an ablation study fashion to evaluate the contribution of each module to the performance (see Section \ref{sec:ablationstudy}).
For instance, the comparison between \myexp{5} and \myexp{4} in Figure \ref{fig:boxplots} shows the influence of the multi-resolution module.
% Similarly, the impact of the data augmentation module can be seen by comparing \myexp{4} and \myexp{3}.
Figure \ref{subfig:boxplot_parenchyma_noabn} shows that all of our experiments outperformed the \secondobs when segmenting the kidney parenchyma in dataset B$_{20}$ (patients without kidney abnormalities).
While the presence of kidney abnormalities affected the performance of kidney (parenchyma + abnormalities) segmentation; see the difference of outliers between Figure \ref{subfig:boxplot_parenchyma_noabn} (dataset B$_{20}$) and Figure \ref{subfig:boxplot_abnparenchyma_abn} (dataset B$_{30}$: patients with kidney abnormalities).
One of the reasons for this behavior may be the difficulty in defining the boundary between the parenchyma and the kidney abnormality.
When comparing the boxplots, the interquartile range of \myExp{5} and \myexp{2} obtained the largest and the smallest interquartile range, respectively, indicating that the combination of multi-resolution, data augmentation, and top-$k$ modules positively impacted the segmentation of the kidneys (parenchyma + abnormalities).
Note the spatial dropout module (difference between \myexp{1} and \myexp{2}) was beneficial only to the kidney abnormality class (see Figure \ref{fig:boxplots}).
Furthermore, Figure \ref{subfig:boxplot_abn_abn} shows that the mean Dice score (black dashed line in boxplots) of our experiments gradually increases when adding more modules (\myexp{5} to \myexp{1}) when segmenting the kidney abnormality class.
This highlights the positive impact of each module in this ablation study on the segmentation of kidney abnormalities.

Additionally, we trained nnUNet, a state-of-the-art segmentation method, on our data and obtained results that were consistent with our previous experiments, except for the kidney abnormality class where nnUNet achieved a 0.521 Dice score compared to 0.488 obtained by \myexp{1}.
To explore further improvements, we combined nnUNet predictions with our best-performing experiments, resulting in an ensemble nnUNet + \myexp{1} that achieved 0.526 Dice score for the kidney abnormality class.
Since nnUNet uses only thresholding as postprocessing, we investigated whether postprocessing nnUNet predictions with our dedicated postprocessing could result in better performance.
This additional postprocessing yielded a 0.585 Dice score, an improvement of +0.064 compared to the original nnUNet with 0.521 Dice score.
While nnUNet is a state-of-the-art segmentation method, our dedicated postprocessing method contributed to further improvement in discarding false positive regions.

We note that the performance of the \secondobs is substantially better than any of our experiments when segmenting only the kidney abnormalities, with an average 0.664 Dice score.
Figure \ref{subfig:boxplot_abn_abn} shows four outliers for the \secondobs, three of these cases obtained a Dice score of zero and one case 0.207.
The volume of these four outliers is 29, 197, 282, and 5769 mm$^3$, three of them are below the median kidney abnormality volume in dataset B$_{30}$ (1421 mm$^3$).
This demonstrates the difficulty of kidney abnormality segmentation, even for experienced radiologists.
The fact that we annotated multiple classes of kidney abnormalities (e.g. tumors, cysts, lesions, and masses) as a single class and the diverse patient anatomy in patients with kidney abnormalities may have contributed to the gap in performance.
\begin{figure}[!ht]
% Figure generated using kidney_abnormality_overlays_paper.mlab
  \centering\footnotesize
% Header
\begin{subfigure}[b]{\textwidth}
\begin{subfigure}[b]{0.03\textwidth}\centering $\mbox{}$\end{subfigure}
    \begin{subfigure}[b]{0.31\textwidth}\centering \textbf{Case 1:} 0.346 Dice score\end{subfigure}
    \begin{subfigure}[b]{0.31\textwidth}\centering \textbf{Case 2:} 0.266 Dice score\end{subfigure}
    \begin{subfigure}[b]{0.31\textwidth}\centering \textbf{Case 3:} 0.777 Dice score\end{subfigure}
\end{subfigure}

% First row of images (Input slice)
\begin{subfigure}[b]{\textwidth}
\vspace{0.2cm}
    \begin{subfigure}[b]{0.03\textwidth}
    \rotatebox[origin=l]{90}{\hspace{1cm}\textbf{(a)} Input slice}
    \end{subfigure}
    %\begin{subfigure}[b]{0.31\textwidth}\includegraphics[width=3.7cm]{Chapter_04_KidneyAbnormality/imgs/predabn/{1.2.392.200036.9116.2.6.1.37.2423319636.1428542387.290431_sag_empty}.png}\end{subfigure}
	\begin{subfigure}[b]{0.31\textwidth}\includegraphics[width=3.7cm]{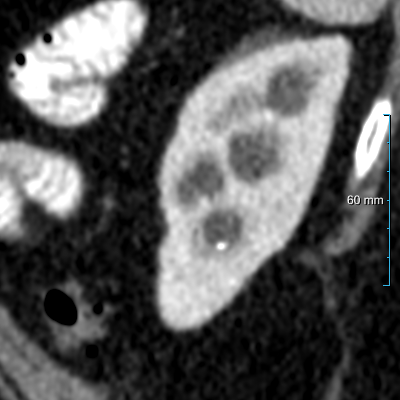}\end{subfigure}
    %\begin{subfigure}[b]{0.31\textwidth}\includegraphics[width=3.7cm]{Chapter_04_KidneyAbnormality/imgs/predabn/{1.3.12.2.1107.5.1.4.51618.30000015091106445600000007968_axi_empty}.png}\end{subfigure}
	\begin{subfigure}[b]{0.31\textwidth}\includegraphics[width=3.7cm]{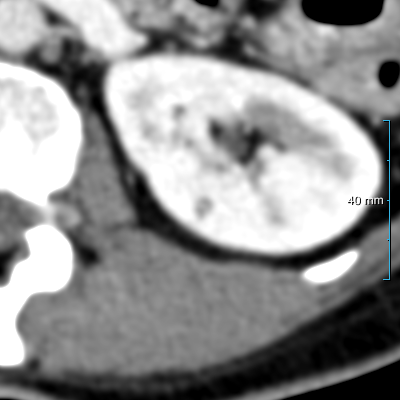}\end{subfigure}
    %\begin{subfigure}[b]{0.31\textwidth}\includegraphics[width=3.7cm]{Chapter_04_KidneyAbnormality/imgs/predabn/{1.3.12.2.1107.5.1.4.54105.30000015081006471059300006697_axi_empty}.png}\end{subfigure}
	\begin{subfigure}[b]{0.31\textwidth}\includegraphics[width=3.7cm]{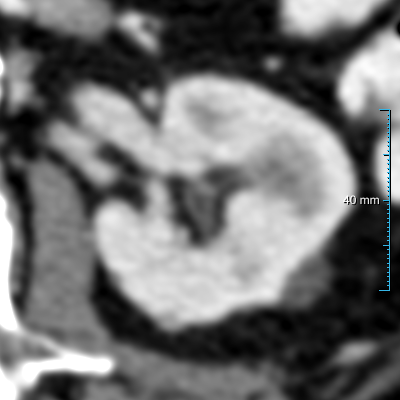}\end{subfigure}
\end{subfigure}

% Second row of images (Heatmaps)
\begin{subfigure}[b]{\textwidth}
    \begin{subfigure}[b]{0.03\textwidth}
    \rotatebox[origin=l]{90}{\hspace{-0.3cm}\textbf{(b)} Pred. of \myexp{1}}
    \end{subfigure}
    %\begin{subfigure}[b]{0.31\textwidth}\includegraphics[width=3.7cm]{Chapter_04_KidneyAbnormality/imgs/predabn/{1.2.392.200036.9116.2.6.1.37.2423319636.1428542387.290431_sag_heatmap}.png}\end{subfigure}
	\begin{subfigure}[b]{0.31\textwidth}\includegraphics[width=3.7cm]{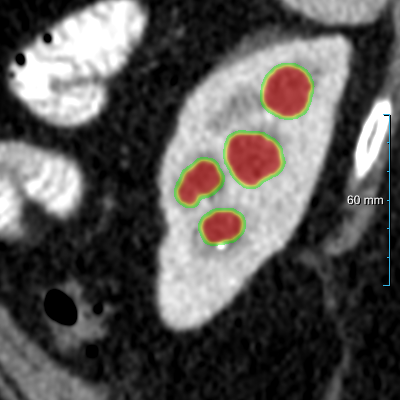}\end{subfigure}
    %\begin{subfigure}[b]{0.31\textwidth}\includegraphics[width=3.7cm]{Chapter_04_KidneyAbnormality/imgs/predabn/{1.3.12.2.1107.5.1.4.51618.30000015091106445600000007968_axi_heatmap}.png}\end{subfigure}
	\begin{subfigure}[b]{0.31\textwidth}\includegraphics[width=3.7cm]{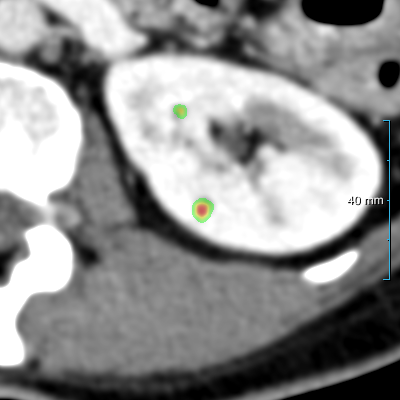}\end{subfigure}
    %\begin{subfigure}[b]{0.31\textwidth}\includegraphics[width=3.7cm]{Chapter_04_KidneyAbnormality/imgs/predabn/{1.3.12.2.1107.5.1.4.54105.30000015081006471059300006697_axi_heatmap}.png}\end{subfigure}
	\begin{subfigure}[b]{0.31\textwidth}\includegraphics[width=3.7cm]{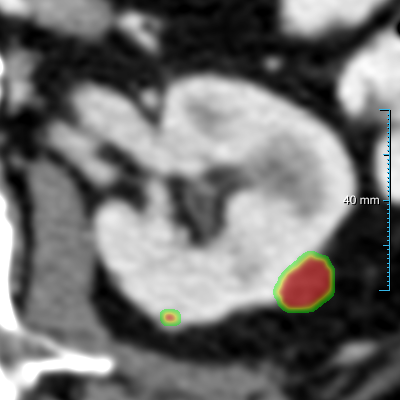}\end{subfigure}
\end{subfigure}

% Third row of images (Contours)
\begin{subfigure}[b]{\textwidth}
    \begin{subfigure}[b]{0.03\textwidth}
    \rotatebox[origin=l]{90}{\hspace{1cm}\textbf{(c)} Contours}
    \end{subfigure}
    %\begin{subfigure}[b]{0.31\textwidth}\includegraphics[width=3.7cm]{Chapter_04_KidneyAbnormality/imgs/predabn/{1.2.392.200036.9116.2.6.1.37.2423319636.1428542387.290431_sag_contours}.png}\end{subfigure}
	\begin{subfigure}[b]{0.31\textwidth}\includegraphics[width=3.7cm]{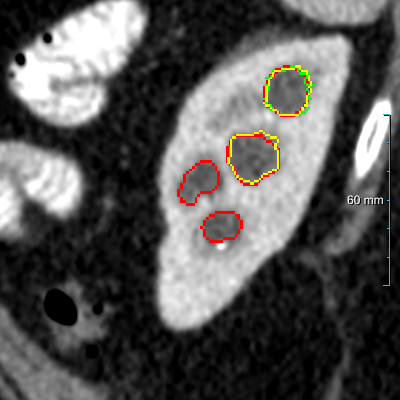}\end{subfigure}
    %\begin{subfigure}[b]{0.31\textwidth}\includegraphics[width=3.7cm]{Chapter_04_KidneyAbnormality/imgs/predabn/{1.3.12.2.1107.5.1.4.51618.30000015091106445600000007968_axi_contours}.png}\end{subfigure}
	\begin{subfigure}[b]{0.31\textwidth}\includegraphics[width=3.7cm]{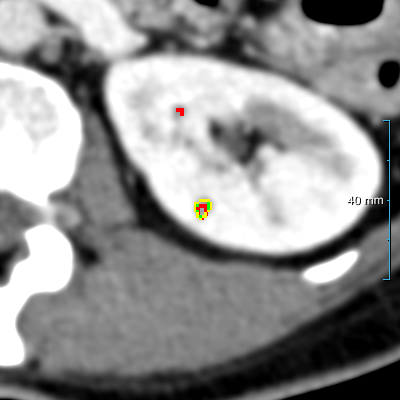}\end{subfigure}
    %\begin{subfigure}[b]{0.31\textwidth}\includegraphics[width=3.7cm]{Chapter_04_KidneyAbnormality/imgs/predabn/{1.3.12.2.1107.5.1.4.54105.30000015081006471059300006697_axi_contours}.png}\end{subfigure}
	\begin{subfigure}[b]{0.31\textwidth}\includegraphics[width=3.7cm]{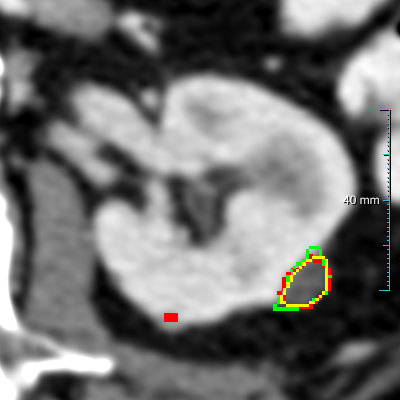}\end{subfigure}
\end{subfigure}

  \caption{Comparison of three cases on the test set B$_{30}$ between \myexp{1}, the reference standard, and the second observer. (a) shows the original slice. (b) shows the heatmaps (predictions prior to post-processing, using a color table mapping [0,1] from transparent to green to red) of \myexp{1}. (c) shows the final predictions (red contours) of \myexp{1}, the reference standard (green contours), and the second human observer (yellow contours). The window center and window width used for all slices were 60 HU and 360 HU.}
  \label{fig:heatmapskidneyabn}
\end{figure}

Table \ref{tab:results_literature} compares the Dice score obtained by previous work and our methods; the middle line separates methods that segmented kidneys without abnormalities and kidneys with abnormalities. 
While some methods reported Dice score for both kidneys as a single score as reported in this paper, others reported Dice scores for the left and right kidneys separately; then, we postprocessed our predictions to the same format and have a better comparison.
Most of the methods trained without kidney abnormalities achieved higher Dice scores in the kidney parenchyma than those trained with kidney abnormalities (below the middle line) due to the more complex task.
Although the performance of \myexp{1} for kidney abnormality segmentation was the lowest (0.487) among the previous work, the performance of the \secondobs (0.664) was also below the previous work where \citet{Yu19} obtained 0.913 and \citet{Yang18} 0.802 Dice score.
This disparity could be due to the fact that we grouped different types of kidney abnormalities including cysts, lesions, masses, metastases, and tumors into a single class while \citet{Yu19} and \citet{Yang18} discarded other abnormalities different than kidney tumors.
Our set of kidney abnormalities is diverse in terms of volume, texture, image intensity, and location in the kidney, which makes network learning difficult.

%We analyzed the volume of the kidney abnormalities in dataset B$_{30}$.
%The minimum, first quartile, median, third quartile, and maximum volume are 29, 204, 1421, 10487, and 639927 mm$^3$, respectively.
%The volume of the smallest renal tumor in the test set of \cite{Yang18} is 6ml and obtained 0.44 Dice score in that case.

\begin{figure}[!ht]
% Figure generated using kidney_abnormality_overlays_paper.mlab
  \centering\scriptsize
% Header
\begin{subfigure}[b]{\textwidth}
    \begin{subfigure}[b]{0.03\textwidth}\centering $\mbox{}$\end{subfigure}
    \begin{subfigure}[b]{0.231\textwidth}\centering \textbf{Case 1} (0.971 Dice) best from dataset $B_{20}$\end{subfigure}
    \begin{subfigure}[b]{0.231\textwidth}\centering \textbf{Case 2} (0.950 Dice) median from dataset $B_{20}$\end{subfigure}
    \begin{subfigure}[b]{0.231\textwidth}\centering \textbf{Case 3} (0.966 Dice) best from dataset $B_{30}$\end{subfigure}
    \begin{subfigure}[b]{0.231\textwidth}\centering \textbf{Case 4} (0.953 Dice) median from dataset $B_{30}$\end{subfigure}
\end{subfigure}

% First row of images (Input slice)
\begin{subfigure}[b]{\textwidth}
\vspace{0.2cm}
    \begin{subfigure}[b]{0.03\textwidth}
    \rotatebox[origin=l]{90}{\hspace{1cm}\textbf{(a)} Input slice}
    \end{subfigure}
    %\begin{subfigure}[b]{0.235\textwidth}\includegraphics[width=2.8cm]{Chapter_04_KidneyAbnormality/imgs/predparenchyma/{1.3.12.2.1107.5.1.4.54105.30000015121307020023400024987_cor_empty_org1}.png}\end{subfigure}
	\begin{subfigure}[b]{0.235\textwidth}\includegraphics[width=2.8cm]{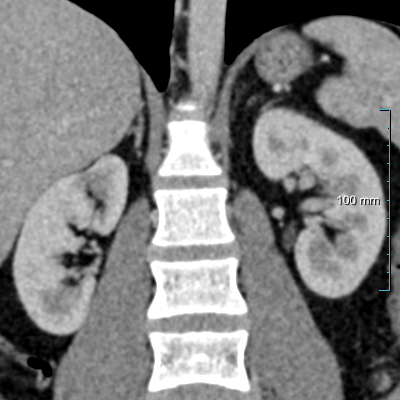}\end{subfigure}
    %\begin{subfigure}[b]{0.235\textwidth}\includegraphics[width=2.8cm]{Chapter_04_KidneyAbnormality/imgs/predparenchyma/{1.3.12.2.1107.5.1.4.51618.30000015061106503951500011840_cor_empty_org1}.png}\end{subfigure}
	\begin{subfigure}[b]{0.235\textwidth}\includegraphics[width=2.8cm]{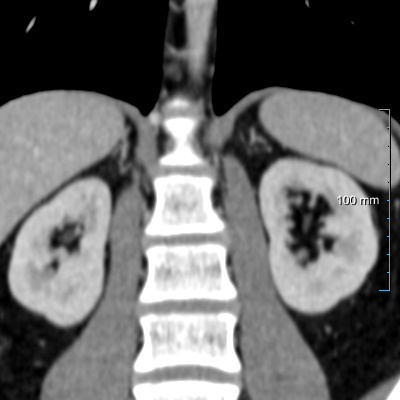}\end{subfigure}
    %\begin{subfigure}[b]{0.235\textwidth}\includegraphics[width=2.8cm]{Chapter_04_KidneyAbnormality/imgs/predparenchyma/{1.3.12.2.1107.5.1.4.54105.30000015081006471059300006697_cor_empty_org1}.png}\end{subfigure}
	\begin{subfigure}[b]{0.235\textwidth}\includegraphics[width=2.8cm]{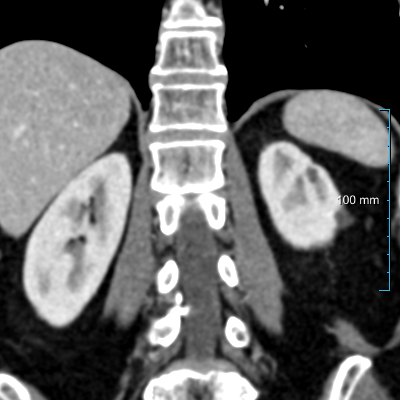}\end{subfigure}
    %\begin{subfigure}[b]{0.235\textwidth}\includegraphics[width=2.8cm]{Chapter_04_KidneyAbnormality/imgs/predparenchyma/{1.3.12.2.1107.5.1.4.51618.30000015111607263893700013851_cor_empty_org1}.png}\end{subfigure}
	\begin{subfigure}[b]{0.235\textwidth}\includegraphics[width=2.8cm]{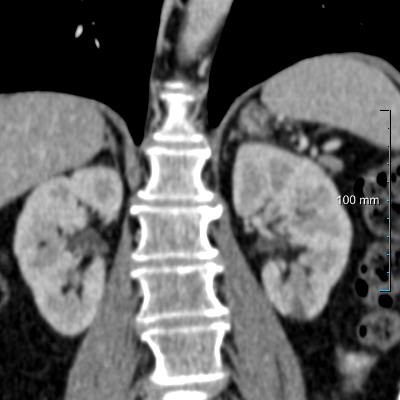}\end{subfigure}
\end{subfigure}

% Second row of images (Contours)
\begin{subfigure}[b]{\textwidth}
    \begin{subfigure}[b]{0.03\textwidth}
    \rotatebox[origin=l]{90}{\hspace{1cm}\textbf{(b)} Contours}
    \end{subfigure}
    %\begin{subfigure}[b]{0.235\textwidth}\includegraphics[width=2.8cm]{Chapter_04_KidneyAbnormality/imgs/predparenchyma/{1.3.12.2.1107.5.1.4.54105.30000015121307020023400024987_cor_contours_org1}.png}\end{subfigure}
	\begin{subfigure}[b]{0.235\textwidth}\includegraphics[width=2.8cm]{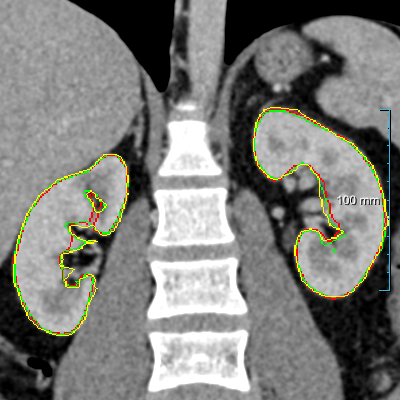}\end{subfigure}
    %\begin{subfigure}[b]{0.235\textwidth}\includegraphics[width=2.8cm]{Chapter_04_KidneyAbnormality/imgs/predparenchyma/{1.3.12.2.1107.5.1.4.51618.30000015061106503951500011840_cor_contours_org1}.png}\end{subfigure}
	\begin{subfigure}[b]{0.235\textwidth}\includegraphics[width=2.8cm]{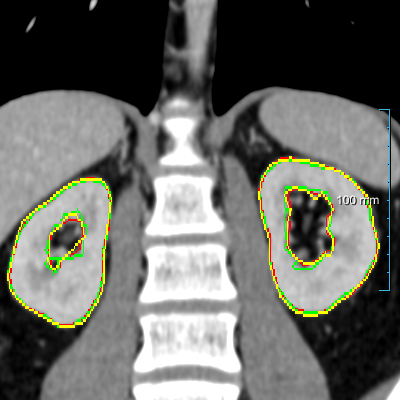}\end{subfigure}
    %\begin{subfigure}[b]{0.235\textwidth}\includegraphics[width=2.8cm]{Chapter_04_KidneyAbnormality/imgs/predparenchyma/{1.3.12.2.1107.5.1.4.54105.30000015081006471059300006697_cor_contours_org1}.png}\end{subfigure}
	\begin{subfigure}[b]{0.235\textwidth}\includegraphics[width=2.8cm]{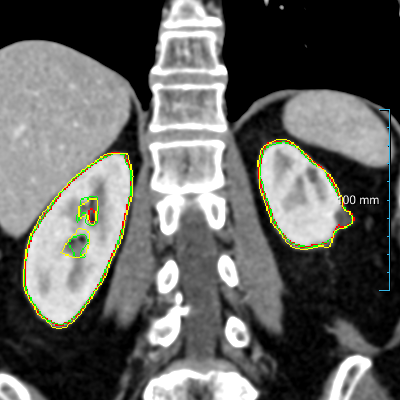}\end{subfigure}
    %\begin{subfigure}[b]{0.235\textwidth}\includegraphics[width=2.8cm]{Chapter_04_KidneyAbnormality/imgs/predparenchyma/{1.3.12.2.1107.5.1.4.51618.30000015111607263893700013851_cor_contours_org1}.png}\end{subfigure}
	\begin{subfigure}[b]{0.235\textwidth}\includegraphics[width=2.8cm]{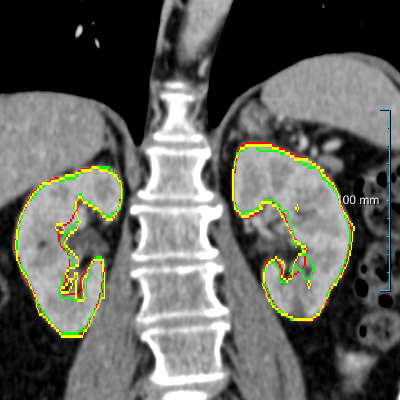}\end{subfigure}
\end{subfigure}
  \caption{Comparison of four cases between \myexp{1}, the reference standard, and the second observer on the test set B$_{30}$ in annotation format 1. (a) shows the original slice and (b) shows the final predictions (red contours) of \myexp{1}, the reference standard (green contours), and the second human observer (yellow contours). All the slices have a window center of 60 HU and a window width of 360 HU.}
  \label{fig:predparenchyma}
\end{figure}

Segmenting kidney abnormalities is challenging due to the similarity between tumors in the collecting system and kidney cysts.
For instance, Figure \ref{fig:heatmapskidneyabn} shows three cases from dataset B$_{30}$ where our method returned some false positives due to the similarity with tumors in the collecting system.
Each case shows the kidney abnormality predictions of \myexp{1} prior to post-processing in the second row as heatmaps.
While the third row shows the post-processed segmentation, reference standard, and second observer as red, green, and yellow contours, respectively.
In all three cases, a false positive by our method is present, indicated by an isolated red contour.
In case 1, the false positives are abnormalities in the collecting system, which have a similar image intensity as the cysts, similarly, the second observer also segmented one of these abnormalities in the middle region.
In case 2, the false positive appears as a small cyst-like region, while in case 3, it resembles an irregular region in the kidney.
Figure \ref{fig:predparenchyma} shows a comparison of the final prediction in annotation format 1 of \myexp{1}, the reference standard, and the second observer represented as red, green, and yellow contours, respectively.
This figure shows the best and median cases of datasets B$_{20}$ and B$_{30}$ and the Dice score of each case computed between \myexp{1} and the reference standard.

A limitation of our study is that we excluded patients with unusual anatomy and with abnormalities in the collecting system.  

%%%%%%%%%%%%%%%%%%%%%%%%%%%%%%%%%%%%%%%%%%%%%%%%%%%%%%%%%%%%%%%%%%%%%%%%%%
%%%%%%%%%%%%%%%%%%%%%%%%%%%%%%%%%%%%%%%%%%%%%%%%%%%%%%%%%%%%%%%%%%%%%%%%%%
\section{Conclusions}
In conclusion, our ablation study and nnUNet showed that segmenting kidney abnormalities in challenging scenarios is possible, and improved performance can be achieved by an ensemble of different methods and dedicated postprocessing.
The results show that our method has the potential to be a valuable tool for clinicians in detecting and monitoring kidney abnormalities.
An ablation study was conducted to better understand the impact of the different modules of our method on its performance.
Further research is needed to optimize the performance of \myexp{1} and nnUNet to test their ability to generalize to other datasets.
Overall, our work contributes to the ongoing efforts to develop accurate and reliable computer-aided diagnosis systems for detecting and quantifying renal abnormalities.

%Bibliography
\bibliographystyle{unsrtnat}
%\bibliography{./Bibliography/fullstrings,./Bibliography/diag,./Bibliography/diagnoweb}
%\bibliography{arxiv_template}

\begin{thebibliography}{40}
\providecommand{\natexlab}[1]{#1}
\providecommand{\url}[1]{\texttt{#1}}
\expandafter\ifx\csname urlstyle\endcsname\relax
  \providecommand{\doi}[1]{doi: #1}\else
  \providecommand{\doi}{doi: \begingroup \urlstyle{rm}\Url}\fi

\bibitem[Ali et~al.(2007)Ali, Farag, and El-Baz]{Ali07}
A.~M. Ali, A.~A. Farag, and A.~S. El-Baz.
\newblock Graph cuts framework for kidney segmentation with prior shape
  constraints.
\newblock In \emph{Medical Image Computing and Computer-Assisted Intervention},
  pages 384--392. Springer, 2007.

\bibitem[Badura and Wieclawek(2016)]{Badu16a}
P.~Badura and W.~Wieclawek.
\newblock Calibrating level set approach by granular computing in computed
  tomography abdominal organs segmentation.
\newblock \emph{Applied Soft Computing}, 49:\penalty0 887--900, 2016.

\bibitem[Berrada et~al.(2018)Berrada, Zisserman, and Kumar]{Berr18a}
L.~Berrada, A.~Zisserman, and M.~P. Kumar.
\newblock Smooth loss functions for deep top-k classification.
\newblock \emph{arXiv preprint arXiv:1802.07595}, 2018.

\bibitem[Bilic et~al.(2023)Bilic, Christ, Li, Vorontsov, Ben-Cohen, Kaissis,
  Szeskin, Jacobs, Humpire Mamani, Chartrand, Lohöfer, Holch, Sommer, Hofmann,
  Hostettler, Lev-Cohain, Drozdzal, Amitai, Vivanti, Sosna, Ezhov, Sekuboyina,
  Navarro, Kofler, Paetzold, Shit, Hu, Lipková, Rempfler, Piraud, Kirschke,
  Wiestler, Zhang, Hülsemeyer, Beetz, Ettlinger, Antonelli, Bae, Bellver, Bi,
  Chen, Chlebus, Dam, Dou, Fu, Georgescu, i~Nieto, Gruen, Han, Heng, Hesser,
  Moltz, Igel, Isensee, Jäger, Jia, Kaluva, Khened, Kim, Kim, Kim, Kohl,
  Konopczynski, Kori, Krishnamurthi, Li, Li, Li, Li, Lowengrub, Ma, Maier-Hein,
  Maninis, Meine, Merhof, Pai, Perslev, Petersen, Pont-Tuset, Qi, Qi, Rippel,
  Roth, Sarasua, Schenk, Shen, Torres, Wachinger, Wang, Weninger, Wu, Xu, Yang,
  Yu, Yuan, Yue, Zhang, Cardoso, Bakas, Braren, Heinemann, Pal, Tang, Kadoury,
  Soler, {van Ginneken}, Greenspan, Joskowicz, and Menze]{Bili23}
P.~Bilic, P.~Christ, H.~B. Li, E.~Vorontsov, A.~Ben-Cohen, G.~Kaissis,
  A.~Szeskin, C.~Jacobs, G.~E.~Humpire Mamani, G.~Chartrand, F.~Lohöfer, J.~W.
  Holch, W.~Sommer, F.~Hofmann, A.~Hostettler, N.~Lev-Cohain, M.~Drozdzal,
  M.~M. Amitai, R.~Vivanti, J.~Sosna, I.~Ezhov, A.~Sekuboyina, F.~Navarro,
  F.~Kofler, J.~C. Paetzold, S.~Shit, X.~Hu, J.~Lipková, M.~Rempfler,
  M.~Piraud, J.~Kirschke, B.~Wiestler, Z.~Zhang, C.~Hülsemeyer, M.~Beetz,
  F.~Ettlinger, M.~Antonelli, W.~Bae, M.~Bellver, L.~Bi, H.~Chen, G.~Chlebus,
  E.~B. Dam, Q.~Dou, C.-W. Fu, B.~Georgescu, X.~G. i~Nieto, F.~Gruen, X.~Han,
  P.-A. Heng, J.~Hesser, J.~H. Moltz, C.~Igel, F.~Isensee, P.~Jäger, F.~Jia,
  K.~C. Kaluva, M.~Khened, I.~Kim, J.-H. Kim, S.~Kim, S.~Kohl, T.~Konopczynski,
  A.~Kori, G.~Krishnamurthi, F.~Li, H.~Li, J.~Li, X.~Li, J.~Lowengrub, J.~Ma,
  K.~Maier-Hein, K.-K. Maninis, H.~Meine, D.~Merhof, A.~Pai, M.~Perslev,
  J.~Petersen, J.~Pont-Tuset, J.~Qi, X.~Qi, O.~Rippel, K.~Roth, I.~Sarasua,
  A.~Schenk, Z.~Shen, J.~Torres, C.~Wachinger, C.~Wang, L.~Weninger, J.~Wu,
  D.~Xu, X.~Yang, S.~C.-H. Yu, Y.~Yuan, M.~Yue, L.~Zhang, J.~Cardoso, S.~Bakas,
  R.~Braren, V.~Heinemann, C.~Pal, A.~Tang, S.~Kadoury, L.~Soler, B.~{van
  Ginneken}, H.~Greenspan, L.~Joskowicz, and B.~Menze.
\newblock The {L}iver {T}umor {S}egmentation {B}enchmark ({L}i{TS}).
\newblock \emph{Medical Image Analysis}, 84:\penalty0 102680, 2023.
\newblock ISSN 1361-8415.
\newblock \doi{10.1016/j.media.2022.102680}.

\bibitem[Blau et~al.(2018)Blau, Klang, Kiryati, Amitai, Portnoy, and
  Mayer]{Blau18}
N.~Blau, E.~Klang, N.~Kiryati, M.~Amitai, O.~Portnoy, and A.~Mayer.
\newblock Fully automatic detection of renal cysts in abdominal {CT} scans.
\newblock \emph{International Journal of Computer Assisted Radiology and
  Surgery}, 13\penalty0 (7):\penalty0 957--966, 2018.

\bibitem[Chen et~al.(2010)Chen, Summers, and Yao]{Chen10g}
X.~Chen, R.~Summers, and J.~Yao.
\newblock {FEM} based 3{D} tumor growth prediction for kidney tumor.
\newblock In \emph{International Workshop on Medical Imaging and Virtual
  Reality}, pages 159--168. Springer, 2010.

\bibitem[Chlebus et~al.(2018)Chlebus, Schenk, Moltz, van Ginneken, Hahn, and
  Meine]{Chle18}
G.~Chlebus, A.~Schenk, J.~H. Moltz, B.~van Ginneken, H.~K. Hahn, and H.~Meine.
\newblock Automatic liver tumor segmentation in {CT} with fully convolutional
  neural networks and object-based postprocessing.
\newblock \emph{Nature Scientific Reports}, 8:\penalty0 15497, Oct. 2018.
\newblock \doi{10.1038/s41598-018-33860-7}.

\bibitem[{\c{C}}i{\c{c}}ek et~al.(2016){\c{C}}i{\c{c}}ek, Abdulkadir, Lienkamp,
  Brox, and Ronneberger]{Cice16}
{\"O}.~{\c{C}}i{\c{c}}ek, A.~Abdulkadir, S.~S. Lienkamp, T.~Brox, and
  O.~Ronneberger.
\newblock 3{D} {U}-{N}et: Learning dense volumetric segmentation from sparse
  annotation.
\newblock In S.~Ourselin, L.~Joskowicz, M.~R. Sabuncu, G.~Unal, and W.~Wells,
  editors, \emph{Medical Image Computing and Computer-Assisted Intervention},
  pages 424--432, Cham, 2016. Springer International Publishing.
\newblock \doi{10.1007/978-3-319-46723-8_49}.
\newblock URL \url{https://doi.org/10.1007/978-3-319-46723-8_49}.

\bibitem[Farmaki et~al.(2010)Farmaki, Marias, Sakkalis, and Graf]{Farm10}
C.~Farmaki, K.~Marias, V.~Sakkalis, and N.~Graf.
\newblock Spatially adaptive active contours: a semi-automatic tumor
  segmentation framework.
\newblock \emph{International Journal of Computer Assisted Radiology and
  Surgery}, 5\penalty0 (4):\penalty0 369--384, 2010.

\bibitem[Gibson et~al.(2018{\natexlab{a}})Gibson, Giganti, Hu, Bonmati,
  Bandula, Gurusamy, Davidson, Pereira, Clarkson, and Barratt]{Gibs18a}
E.~Gibson, F.~Giganti, Y.~Hu, E.~Bonmati, S.~Bandula, K.~Gurusamy, B.~Davidson,
  S.~P. Pereira, M.~J. Clarkson, and D.~C. Barratt.
\newblock Automatic multi-organ segmentation on abdominal {CT} with dense
  {V-N}etworks.
\newblock \emph{IEEE Transactions on Medical Imaging}, 37\penalty0
  (8):\penalty0 1822--1834, aug 2018{\natexlab{a}}.
\newblock \doi{10.1109/tmi.2018.2806309}.

\bibitem[Gibson et~al.(2018{\natexlab{b}})Gibson, Li, Sudre, Fidon, Shakir,
  Wang, Eaton-Rosen, Gray, Doel, Hu, et~al.]{Gibs18c}
E.~Gibson, W.~Li, C.~Sudre, L.~Fidon, D.~I. Shakir, G.~Wang, Z.~Eaton-Rosen,
  R.~Gray, T.~Doel, Y.~Hu, et~al.
\newblock Nifty{N}et: a deep-learning platform for medical imaging.
\newblock \emph{Computer Methods and Programs in Biomedicine}, 158:\penalty0
  113--122, 2018{\natexlab{b}}.

\bibitem[Glorot and Bengio(2010)]{Glor10}
X.~Glorot and Y.~Bengio.
\newblock Understanding the difficulty of training deep feedforward neural
  networks.
\newblock In \emph{International conference on artificial intelligence and
  statistics}, pages 249--256, 2010.

\bibitem[{Haghighi} et~al.(2018){Haghighi}, {Warfield}, and {Kurugol}]{Hagh18}
M.~{Haghighi}, S.~K. {Warfield}, and S.~{Kurugol}.
\newblock Automatic renal segmentation in {DCE-MRI} using convolutional neural
  networks.
\newblock In \emph{IEEE International Symposium on Biomedical Imaging}, pages
  1534--1537, Apr. 2018.
\newblock \doi{10.1109/ISBI.2018.8363865}.

\bibitem[Han(2017)]{Han17}
X.~Han.
\newblock Automatic liver lesion segmentation using a deep convolutional neural
  network method.
\newblock \emph{arXiv:1704.07239}, 2017.
\newblock \doi{10.1002/mp.12155}.

\bibitem[Heckel et~al.(2009)Heckel, Schwier, and Peitgen]{Heck09}
F.~Heckel, M.~Schwier, and H.-O. Peitgen.
\newblock Object-oriented application development with {MeVisLab} and {Python}.
\newblock \emph{Lecture Notes in Informatics (Informatik 2009: Im Focus das
  Leben)}, 154:\penalty0 1338--1351, 2009.

\bibitem[Heinrich et~al.(2019)Heinrich, Oktay, and Bouteldja]{Hein19}
M.~P. Heinrich, O.~Oktay, and N.~Bouteldja.
\newblock {OBELISK-N}et: Fewer layers to solve 3{D} multi-organ segmentation
  with sparse deformable convolutions.
\newblock \emph{Medical Image Analysis}, 54:\penalty0 1--9, May 2019.
\newblock ISSN 1361-8423.
\newblock \doi{10.1016/j.media.2019.02.006}.

\bibitem[Heller et~al.(2021)Heller, Isensee, Maier-Hein, Hou, Xie, Li, Nan, Mu,
  Lin, Han, Yao, Gao, Zhang, Wang, Hou, Yang, Xiong, Tian, Zhong, Ma, Rickman,
  Dean, Stai, Tejpaul, Oestreich, Blake, Kaluzniak, Raza, Rosenberg, Moore,
  Walczak, Rengel, Edgerton, Vasdev, Peterson, McSweeney, Peterson, Kalapara,
  Sathianathen, Papanikolopoulos, and Weight]{Hell21}
N.~Heller, F.~Isensee, K.~H. Maier-Hein, X.~Hou, C.~Xie, F.~Li, Y.~Nan, G.~Mu,
  Z.~Lin, M.~Han, G.~Yao, Y.~Gao, Y.~Zhang, Y.~Wang, F.~Hou, J.~Yang, G.~Xiong,
  J.~Tian, C.~Zhong, J.~Ma, J.~Rickman, J.~Dean, B.~Stai, R.~Tejpaul,
  M.~Oestreich, P.~Blake, H.~Kaluzniak, S.~Raza, J.~Rosenberg, K.~Moore,
  E.~Walczak, Z.~Rengel, Z.~Edgerton, R.~Vasdev, M.~Peterson, S.~McSweeney,
  S.~Peterson, A.~Kalapara, N.~Sathianathen, N.~Papanikolopoulos, and
  C.~Weight.
\newblock The state of the art in kidney and kidney tumor segmentation in
  contrast-enhanced {CT} imaging: {R}esults of the {KiTS}19 challenge.
\newblock \emph{Medical Image Analysis}, 67:\penalty0 101821, 2021.

\bibitem[Isensee et~al.(2020)Isensee, Jaeger, Kohl, Petersen, and
  Maier-Hein]{Isen20}
F.~Isensee, P.~F. Jaeger, S.~A.~A. Kohl, J.~Petersen, and K.~H. Maier-Hein.
\newblock nn{U}-{N}et: a self-configuring method for deep learning-based
  biomedical image segmentation.
\newblock \emph{Nature methods}, 2020.
\newblock ISSN 1548-7091.
\newblock \doi{10.1038/s41592-020-01008-z}.

\bibitem[Jackson et~al.(2018)Jackson, Hardcastle, Dawe, Kron, Hofman, and
  Hicks]{Jack18}
P.~Jackson, N.~Hardcastle, N.~Dawe, T.~Kron, M.~Hofman, and R.~J. Hicks.
\newblock Deep learning renal segmentation for fully automated radiation dose
  estimation in unsealed source therapy.
\newblock \emph{Frontiers in oncology}, 8:\penalty0 215, 2018.

\bibitem[Kaur et~al.(2019)Kaur, Juneja, and Mandal]{Kaur19}
R.~Kaur, M.~Juneja, and A.~Mandal.
\newblock A hybrid edge-based technique for segmentation of renal lesions in
  {CT} images.
\newblock \emph{Multimedia Tools and Applications}, 78\penalty0 (10):\penalty0
  12917--12937, 2019.

\bibitem[Khalifa et~al.(2017)Khalifa, Soliman, Elmaghraby, Gimelfarb, and
  El-Baz]{Khal17}
F.~Khalifa, A.~Soliman, A.~Elmaghraby, G.~Gimelfarb, and A.~El-Baz.
\newblock 3{D} kidney segmentation from abdominal images using
  spatial-appearance models.
\newblock \emph{Computational and mathematical methods in medicine}, 2017,
  2017.

\bibitem[Kim and Park(2004)]{Kim04a}
D.-Y. Kim and J.-W. Park.
\newblock Computer-aided detection of kidney tumor on abdominal computed
  tomography scans.
\newblock \emph{Acta Radiologica}, 45\penalty0 (7):\penalty0 791--795, 2004.

\bibitem[Kingma and Ba(2014)]{King15}
D.~Kingma and J.~Ba.
\newblock Adam: {A} method for stochastic optimization.
\newblock \emph{arXiv preprint arXiv:1412.6980}, 2014.

\bibitem[Kutikov and Uzzo(2009)]{Kuti09}
A.~Kutikov and R.~G. Uzzo.
\newblock The {RENAL} nephrometry score: a comprehensive standardized system
  for quantitating renal tumor size, location and depth.
\newblock \emph{Journal of Urology}, 182\penalty0 (3):\penalty0 844--853, 2009.

\bibitem[Lin et~al.(2006)Lin, Lei, and Hung]{Lin06}
D.-T. Lin, C.-C. Lei, and S.-W. Hung.
\newblock Computer-aided kidney segmentation on abdominal {CT} images.
\newblock \emph{IEEE Transactions on Information Technology in Biomedicine},
  10\penalty0 (1):\penalty0 59--65, 2006.

\bibitem[Linguraru et~al.(2009)Linguraru, Yao, Gautam, Peterson, Li, Linehan,
  and Summers]{Ling09}
M.~G. Linguraru, J.~Yao, R.~Gautam, J.~Peterson, Z.~Li, W.~M. Linehan, and
  R.~M. Summers.
\newblock Renal tumor quantification and classification in contrast-enhanced
  abdominal {CT}.
\newblock \emph{Pattern Recognition}, 42\penalty0 (6):\penalty0 1149--1161,
  2009.

\bibitem[Lu et~al.(2007)Lu, Chen, Zhang, and Yang]{Lu07}
J.~Lu, J.~Chen, J.~Zhang, and W.~Yang.
\newblock Segmentation of kidney using {CV} model and anatomy priors.
\newblock In \emph{Medical Imaging, Parallel Processing of Images, and
  Optimization Techniques}, volume 6789 of \emph{Proceedings of the SPIE}, page
  678911, 2007.

\bibitem[Ronneberger et~al.(2015)Ronneberger, Fischer, and Brox]{Ronn15}
O.~Ronneberger, P.~Fischer, and T.~Brox.
\newblock U-{N}et: Convolutional networks for biomedical image segmentation.
\newblock In \emph{Medical Image Computing and Computer-Assisted Intervention},
  volume 9351 of \emph{Lecture Notes in Computer Science}, pages 234--241,
  2015.

\bibitem[Sharma et~al.(2017)Sharma, Rupprecht, Caroli, Aparicio, Remuzzi,
  Baust, and Navab]{Shar17}
K.~Sharma, C.~Rupprecht, A.~Caroli, M.~C. Aparicio, A.~Remuzzi, M.~Baust, and
  N.~Navab.
\newblock Automatic segmentation of kidneys using deep learning for total
  kidney volume quantification in autosomal dominant polycystic kidney disease.
\newblock \emph{Nature Scientific Reports}, 7\penalty0 (1):\penalty0 2049,
  2017.
\newblock \doi{10.1038/s41598-017-01779-0}.

\bibitem[Siegel et~al.(2019)Siegel, Miller, and Jemal]{Sieg19}
R.~L. Siegel, K.~D. Miller, and A.~Jemal.
\newblock Cancer statistics, 2019.
\newblock \emph{CA: A Cancer Journal for Clinicians}, 69\penalty0 (1):\penalty0
  7--34, 2019.

\bibitem[Skalski et~al.(2017)Skalski, Heryan, Jakubowski, and Drewniak]{Skal17}
A.~Skalski, K.~Heryan, J.~Jakubowski, and T.~Drewniak.
\newblock Kidney segmentation in {CT} data using hybrid {L}evel-{S}et method
  with ellipsoidal shape constraints.
\newblock \emph{Metrology and Measurement Systems}, 24\penalty0 (1):\penalty0
  101--112, 2017.
\newblock \doi{10.1515/mms-2017-0006}.

\bibitem[Taha et~al.(2018)Taha, Lo, Li, and Zhao]{Taha18}
A.~Taha, P.~Lo, J.~Li, and T.~Zhao.
\newblock Kid-{N}et: convolution networks for kidney vessels segmentation from
  {CT}-volumes.
\newblock In \emph{Medical Image Computing and Computer-Assisted Intervention},
  pages 463--471. Springer, 2018.

\bibitem[Tompson et~al.(2015)Tompson, Goroshin, Jain, LeCun, and
  Bregler]{Tomp15}
J.~Tompson, R.~Goroshin, A.~Jain, Y.~LeCun, and C.~Bregler.
\newblock Efficient object localization using convolutional networks.
\newblock In \emph{Computer Vision and Pattern Recognition}, pages 648--656,
  2015.

\bibitem[Turco et~al.(2018)Turco, Valinoti, Martin, Tagliaferri, Scolari, and
  Corsi]{Turc18}
D.~Turco, M.~Valinoti, E.~M. Martin, C.~Tagliaferri, F.~Scolari, and C.~Corsi.
\newblock Fully automated segmentation of polycystic kidneys from noncontrast
  computed tomography: A feasibility study and preliminary results.
\newblock \emph{Academic Radiology}, 25\penalty0 (7):\penalty0 850--855, 2018.

\bibitem[Wang et~al.(2018)Wang, Zhou, Tang, Shen, Fishman, and Yuille]{Wang18c}
Y.~Wang, Y.~Zhou, P.~Tang, W.~Shen, E.~K. Fishman, and A.~L. Yuille.
\newblock Training multi-organ segmentation networks with sample selection by
  relaxed upper confident bound.
\newblock In \emph{Medical Image Computing and Computer-Assisted Intervention},
  pages 434--442. Springer, 2018.

\bibitem[Wieclawek(2018)]{Wiec18}
W.~Wieclawek.
\newblock 3{D} marker-controlled watershed for kidney segmentation in clinical
  {CT} exams.
\newblock \emph{Biomedical Engineering Online}, 17\penalty0 (1):\penalty0 26,
  2018.

\bibitem[{Yang} et~al.(2018){Yang}, {Li}, {Pan}, {Kong}, {Wu}, {Shu}, {Luo},
  {Dillenseger}, {Coatrieux}, {Tang}, and {Zhu}]{Yang18}
G.~{Yang}, G.~{Li}, T.~{Pan}, Y.~{Kong}, J.~{Wu}, H.~{Shu}, L.~{Luo},
  J.~{Dillenseger}, J.~{Coatrieux}, L.~{Tang}, and X.~{Zhu}.
\newblock Automatic segmentation of kidney and renal tumor in {CT} images based
  on {3D} fully convolutional neural network with pyramid pooling module.
\newblock In \emph{International Conference on Pattern Recognition}, pages
  3790--3795, Aug. 2018.
\newblock \doi{10.1109/ICPR.2018.8545143}.

\bibitem[Yoruk et~al.(2018)Yoruk, Hargreaves, and Vasanawala]{Yoru18}
U.~Yoruk, B.~A. Hargreaves, and S.~S. Vasanawala.
\newblock Automatic renal segmentation for {MR} urography using
  3{D}-{G}rab{C}ut and random forests.
\newblock \emph{Magnetic Resonance in Medicine}, 79\penalty0 (3):\penalty0
  1696--1707, 2018.

\bibitem[{Yu} et~al.(2019){Yu}, {Shi}, {Sun}, {Gao}, {Zhu}, and {Dai}]{Yu19}
Q.~{Yu}, Y.~{Shi}, J.~{Sun}, Y.~{Gao}, J.~{Zhu}, and Y.~{Dai}.
\newblock Crossbar-{N}et: A novel convolutional neural network for kidney tumor
  segmentation in {CT} images.
\newblock \emph{IEEE Transactions on Image Processing}, 28\penalty0
  (8):\penalty0 4060--4074, 2019.
\newblock ISSN 1057-7149.
\newblock \doi{10.1109/TIP.2019.2905537}.

\bibitem[Zheng et~al.(2017)Zheng, Liu, Georgescu, Xu, and Comaniciu]{Zhen17}
Y.~Zheng, D.~Liu, B.~Georgescu, D.~Xu, and D.~Comaniciu.
\newblock Deep learning based automatic segmentation of pathological kidney in
  {CT}: Local versus global image context.
\newblock In \emph{Advances in Computer Vision and Pattern Recognition}, pages
  241--255. Springer, Cham, 2017.
\newblock \doi{10.1007/978-3-319-42999-1_14}.

\end{thebibliography}

\end{document}